\documentclass[%
reprint,
superscriptaddress,
amsart,
amsmath,amssymb,
aps,
]{revtex4-2}

\usepackage{graphicx}
\usepackage{dcolumn}
\usepackage{bm}
\usepackage{comment}
\usepackage{braket}
\usepackage{upgreek}
\usepackage{appendix}
\usepackage{xcolor}
\usepackage{dsfont}
\usepackage{ulem}
\usepackage[colorlinks=true, linkcolor=magenta, urlcolor=magenta, citecolor=magenta]{hyperref}

\begin{document}

\preprint{APS/123-QED}

\title{Spatial correlations in four-wave mixing with structured light}

\author{M. R. L. da Motta}
\affiliation{Departamento de Física, Universidade Federal de Pernambuco, 50670-901, Recife, PE, Brazil}

\author{S. S. Vianna}
\email{sandra.vianna@ufpe.br}
\affiliation{Departamento de Física, Universidade Federal de Pernambuco, 50670-901, Recife, PE, Brazil}




\date{\today}

\begin{abstract}
We present a detailed theoretical treatment of four-wave mixing (FWM) in a quantized paraxial framework, capturing the multi-spatial-mode nature of the biphoton state generated in the process. By analyzing the biphoton state both in position and momentum representations, we identify the conditions under which these descriptions become equivalent.
We also highlight formal and physical similarities between FWM and spontaneous parametric down-conversion (PDC), showing that the transfer of pump structure to the spatial coincidence profile, an important and well-known characteristic of the biphoton state, carries over naturally to FWM. In addition, our treatment captures the transition from position correlations in the near field to momentum correlations in the far field, reflecting the underlying spatial entanglement.
The measures of entanglement, including the spiral bandwidth and the Schmidt rank, are discussed. Our work consolidates known and new results on spatial correlations in FWM and provides a theoretical framework that may support future studies in nonlinear and quantum optics with structured light.
\end{abstract}

\maketitle


\section{Introduction}

In recent years, the spatial structure of light has seen a significant increase in research interest, both in fundamental studies and in applications and technological developments \cite{rubinsztein2016roadmap}.
Investigations on the behavior of light fields carrying nontrivial transverse distributions and the ability to control the spatial degrees of freedom of light beams have allowed numerous advances in optical sciences \cite{forbes2021structured}.
These advances range from areas related to fundamental properties of the electromagnetic field \cite{calvo2006quantum,Ballentine2016} to applications in the manipulation of matter \cite{franke2007optical,padgett2011tweezers,melo2020optical}, information multiplexing \cite{pan2019orbital}, nonlinear optics \cite{Faccio2013,buono2022nonlinear}, and many others \cite{yao2011orbital,bliokh2023roadmap}.

The seminal work of Allen \textit{et al.} in 1992 \cite{allen92} is regarded as the origin of all these advances. In it, the authors establish the connection between the orbital angular momentum (OAM) of a light beam and its spatial distribution.
This breakthrough originated the field of light OAM, which over the past three decades has grown immensely and transformed in such a way as to be recognized today as the more general field of structured light \cite{franke202230}.
Shortly after the initial developments, the investigation of the role played by OAM in nonlinear optical processes began in second harmonic generation \cite{Dholakia:1996,Courtial:1997,Berzanskis:1998}.
Today, SHG and other second-order optical phenomena offer a highly versatile platform for studying the transverse degrees of freedom of light.

Four-wave mixing, a third-order nonlinear optical process, has also been extensively employed in the context of structured light \cite{walker2012trans,akulshin2015distinguishing,akulshin2016arithmetic,offer2018spiral,chopinaud2018high,prajapati2019optical,offer2021gouy}.
In addition, four-wave mixing has also emerged as an important method for generating quantum-correlated beams \cite{boyer2008entangled}.
The quantum theory that describes spatial correlations in four-wave mixing (FWM) in the regime of a classical pump bears great similarity to the theory for parametric down-conversion (PDC), which is already well established \cite{walborn2010spatial}.
In very simple terms, in both processes, a pump beam excites the nonlinear medium, and a pair of correlated photons is generated. The difference is that in PDC, a single photon is absorbed from the pump, while in FWM two photons are absorbed.
Of course, this is a general comparison from a fundamental point of view, and more specific aspects must be considered when studying each process in their many configurations.
Nonetheless, many aspects can be translated almost effortlessly to FWM.

An interesting capability thoroughly explored in PDC is to engineer the spatial correlations between the pair of generated photons by modifying the pump beam angular spectrum \cite{monken1998transfer,walborn2010spatial}.
Despite that, to our knowledge, this approach has only recently been employed to encode information on spatial correlations between FWM twin beams with a high degree of control \cite{nirala2023information}.
The system used to achieve this was a sample of heated Rb atoms, carefully optimized with the objective of observing the spatial characteristics of the generated fields \cite{boyer2008generation,boyer2008entangled}.

In this paper, we address the quantum-mechanical theory of FWM induced by structured light, evidencing the multi-spatial-mode character of the generated light-state and the associated spatial correlations.
The solution to Schrödinger's equation with the relevant nonlinear interaction Hamiltonian leads to the biphoton state, which can be cast in the position and momentum (wave-vector) spaces.
Here, we discuss both descriptions and the connections between them.
Our approach is based on the projection of the biphoton state -- either in the position or momentum spaces -- onto a given basis of paraxial modes.
With this method, it becomes natural to identify the contributions of the various spatial modes which constitute the entangled light-state.
This is also useful from a practical point of view, as it allows one to focus on particular subspaces of modes, simplifying the calculations of the entanglement properties \cite{baghdasaryan2022generalized}.

In this context, we study a number of properties of the biphoton state.
We calculate the spatial coincidence count rate and show that the transfer of the pump angular spectrum to the coincidence count profile, a well-known result in PDC \cite{monken1998transfer}, is also verified in FWM in a particular detection configuration.
As already mentioned, this is an expected result due to the similarity between the quantum-mechanical description of the two processes.
In a different detection scheme, we analyze the evolution of the spatial coincidence profile as the generated photons propagate from the non-linear medium exit to a position of the order of the Rayleigh range.
In this case, we show that the transverse coincidence profile undergoes a rotation between the near- and far-field regions, which is consistent with experimental results obtained in PDC \cite{dehghan2024biphoton}.
We also discuss measures of entanglement of the spatial biphoton state, such as the spiral bandwidth and the Schmidt number, and how these quantities are affected by the system parameters.
Our development also provides expressions for the coincidence amplitudes in the phase-matched and phase-mismatched settings.

The paper is organized as follows. In Section \ref{QPFWM} we develop the FWM interaction Hamiltonian in the second-quantized paraxial framework. Next, in Sections \ref{BPSPOS} and \ref{biphoton_momentum} we analyze the FWM biphoton state in the position and momentum representations, respectively. The properties of the biphoton state in both representations are investigated. We conclude with a summary and give the final remarks in Section \ref{Conclusions}. Additional technical details and supporting derivations are provided in the appendices.


\section{Quantum picture of the FWM process}
\label{QPFWM}

\begin{figure}[bt!]
    \centering
    \includegraphics[width=1\linewidth]{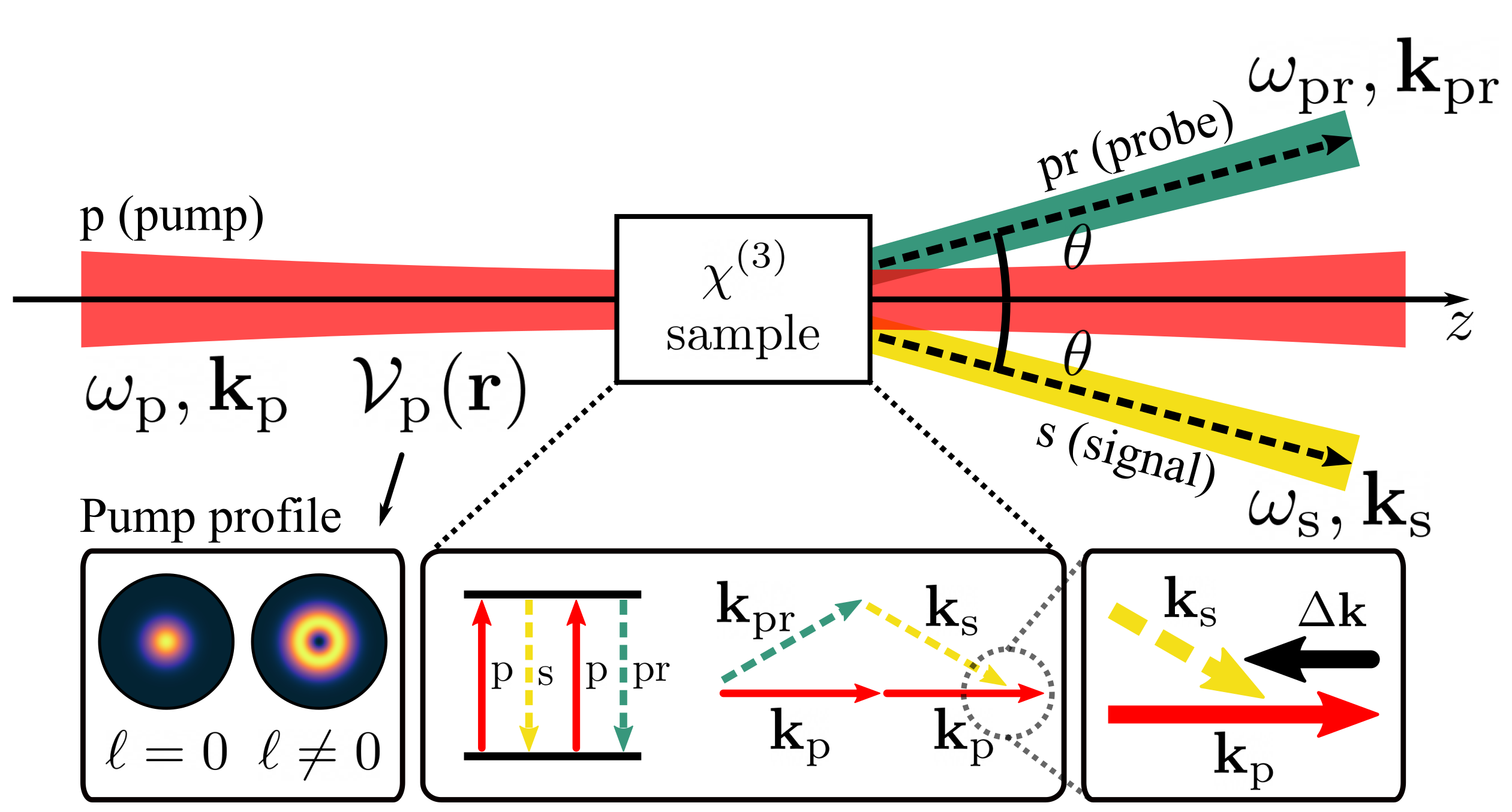}
    \caption{Four-wave mixing configuration. A strong optical pump field with spatial profile $\mathcal{V}_\mathrm{p}(\mathbf{r})$ shines on the third-order nonlinear medium. The pump may carry Gaussian or non-Gaussian spatial distributions, as indicated in the leftmost inset. As a result of the nonlinear interactions, two photons from the pump are absorbed, and two photons are generated on the signal and probe channels. The pair of generated beams are symmetrically distributed, at an angle $\theta$, with respect to the pump propagation direction, which is parallel to the $\mathbf{e}_z$ direction. The central inset represents the conservation of energy and momentum in the process. The rightmost inset illustrates the inherent phase-mismatch associated with the frequency degeneracy condition.}
    \label{fig:sys}
\end{figure}

Consider a sample of atoms centered at the origin and a strong optical field, which we call the pump, labeled $\mathrm{p}$, with frequency $\omega_\mathrm{p}$, wave-vector $\mathbf{k}_\mathrm{p}$ parallel to the $\mathbf{e}_z$ direction, intersecting this third-order nonlinear medium, as indicated in Figure \ref{fig:sys}. As a result of the nonlinear interaction, two photons from the pump are destroyed to generate two photons in the fields we call signal and probe, labeled $\mathrm{s}$ and $\mathrm{pr}$, respectively, with frequencies $(\omega_\mathrm{s},\omega_\mathrm{pr})$ and wave-vectors $(\mathbf{k}_\mathrm{s},\mathbf{k}_\mathrm{pr})$. Due to the energy and momentum conservation requirements, this process must meet the conditions
\begin{align}
    2\omega_\mathrm{p}&=\omega_\mathrm{pr}+\omega_\mathrm{s},
    \\
    2\mathbf{k}_\mathrm{p}&=\mathbf{k}_\mathrm{pr}+\mathbf{k}_\mathrm{s}+\Delta \mathbf{k},
\end{align}
where $\Delta \mathbf{k}$ is the phase-mismatch vector.
Throughout this work, we will consider a degenerate FWM process taking place in a sample of $\mathrm{Rb}^{87}$ atoms, between the $\ket{5\mathrm{S}_{1/2},F=2}$ and $\ket{5\mathrm{P}_{3/2},F=3}$ levels (cyclic transition of the $\mathrm{D}_2$ line), such that the pump wavelength is $\lambda_\mathrm{p}=780\,\mathrm{nm}$ \cite{steck2001rubidium}.

\subsection{The nonlinear interaction Hamiltonian}

For a quantum description of this FWM process, we write the Hamiltonian of our system as $\hat{H} = \hat{H}_0 + \hat{H}_I$, where $\hat{H}_0=\hat{H}_A+\hat{H}_F$ is the unperturbed Hamiltonian, taking into account the atomic ($A$) and light field ($F$) contributions; and $\hat{H}_I$ is the interaction Hamiltonian, given by:
\begin{align}
    \hat{H}_I = \frac{1}{2}\int_\mathbb{V}\left(\hat{\mathbf{E}}^\dagger(\mathbf{r},t)\cdot\hat{\mathbf{P}}(\mathbf{r},t) + \mathrm{h.c.}\right){{d}}^3\mathbf{r},
\end{align}
where $\mathbb{V}$ is the region of space the system occupies, $\hat{\mathbf{E}}$ is the electric field operator, with $(^\dagger)$ denoting the hermitian conjugate (h.c.), and $\hat{\mathbf{P}}$ is the macroscopic polarization operator, which can be separated into its linear and nonlinear components, $\hat{\mathbf{P}}=\hat{\mathbf{P}}_{L}+\hat{\mathbf{P}}_{NL}$.
Our focus is on the nonlinear interaction, which in the case of FWM takes the standard form \cite{nirala2023information}:
\begin{align}
    \hat{H}_I=\frac{\varepsilon_0}{2}\int_\mathbb{V}\left(\chi^{(3)}\hat{E}^\dagger_{\mathrm{s}}\hat{E}_{\mathrm{p}}\hat{E}^\dagger_{\mathrm{pr}}\hat{E}_{\mathrm{p}}+\mathrm{h.c.}\right){{d}}^3\mathbf{r},
\end{align}
where $\chi^{(3)}$ is the third-order nonlinear susceptibility associated with the four-wave mixing process.
We consider that the strong pump $\hat{E}_\mathrm{p}$ can be regarded as a classical monochromatic wave of the form
\begin{align}
    \hat{E}_{\mathrm{p}}(\mathbf{r},t) \rightarrow E_{\mathrm{p}}(\mathbf{r},t) = \mathcal{E}^0_{\mathrm{p}} \mathcal{V}_{\mathrm{p}}(\mathbf{r})e^{-i(k_{\mathrm{p}}z-\omega_{\mathrm{p}}t)},
\end{align}
where $\mathcal{E}^0_{\mathrm{p}}$ is related to the pump power, $k_\mathrm{p}=2\pi/\lambda_\mathrm{p}$, and
\begin{align} \label{eq:Vpump}
    \mathcal{V}_{\mathrm{p}}(\mathbf{r})=\sum_{l,q}c_{l,q}{u}_{l,q}(\mathbf{r})
\end{align}
is the pump spatial distribution, with $\sum_{l,q}|c_{l,q}|^2=1$, and the $u_{l,q}(\mathbf{r})$ functions being the Laguerre-Gaussian paraxial modes, defined in Appendix \ref{LG_modes}.
We shall use the pair of azimuthal and radial indices $\{l,q\}$ to characterize the mode composition of the pump field, in contrast to $\{\ell,p\}$, which will be reserved for the signal and probe channels, with the respective subscripts.

The generated fields, signal $\hat{E}_\mathrm{s}$ and probe $\hat{E}_\mathrm{pr}$, are in turn regarded as operators, and we write them in the general form
\begin{align}
    \hat{E}_j(\mathbf{r},t)=-i\sum_{\mathbf{k}_j}\sqrt{\frac{\hbar\omega_{\mathbf{k}_j}}{2\varepsilon_0V}}\hat{\mathfrak{a}}_{\mathbf{k}_j}e^{-i(\mathbf{k}_j\cdot\mathbf{r}-\omega_{\mathbf{k}_j} t)},
\end{align}
where $j=\mathrm{pr}$ or $\mathrm{s}$ labels the field operator, $V$ is the quantization volume, the wave-vector is $\mathbf{k}_j=(\boldsymbol{\rho},k)_j=\boldsymbol{\rho}_j+\mathbf{e}_zk_j\cos\theta$, $\boldsymbol{\rho}_j$ is the transverse component of the wave-vector, $k_j$ is the longitudinal wave-number, $\theta$ is the separation angle between the pump and signal/probe fields in the experimental configuration, and $\omega_{\mathbf{k}_j}=c|\mathbf{k}_j|\simeq ck_j$ is the frequency.
The bosonic operators satisfy the commutation relation $[\hat{\mathfrak{a}}_\mathbf{k},\hat{\mathfrak{a}}^\dagger_{\mathbf{k}^\prime}]=\delta^{(3)}(\mathbf{k}-\mathbf{k}^\prime)$.
We consider a well-defined polarization for all fields: parallel circular polarization, in a way that we may think of a two-level atom. Therefore, no polarization labels are present.
The nonlinear Hamiltonian, which is our prime interest, can be written as \footnote{It is worth mentioning that the product $\hat{\mathfrak{a}}^\dagger_{\mathbf{k}_{\mathrm{pr}}}\hat{\mathfrak{a}}^\dagger_{\mathbf{k}_{\mathrm{s}}}$ is a tensor product of creation operators acting on different spaces, but for simplicity we omit the symbol $\otimes$.}
\begin{align} \label{eq:H_start}
    \hat{H}_I &= -\frac{\hbar(\mathcal{E}^0_{\mathrm{p}})^2}{4V}\sum_{\mathbf{k}_{\mathrm{pr}},\mathbf{k}_{\mathrm{s}}}\left(\vphantom{\int_\mathbb{V}}\sqrt{\omega_{\mathbf{k}_{\mathrm{pr}}}\omega_{\mathbf{k}_{\mathrm{s}}}}\hat{\mathfrak{a}}^\dagger_{\mathbf{k}_{\mathrm{pr}}}\hat{\mathfrak{a}}^\dagger_{\mathbf{k}_{\mathrm{s}}}e^{i\delta\omega t}\right. \nonumber
    \\
    &\quad\left.\times\int_\mathbb{V}\chi^{(3)}\mathcal{V}_{\mathrm{p}}^2(\mathbf{r})e^{-i\left(2\mathbf{k}_{\mathrm{p}}-\mathbf{k}_{\mathrm{pr}}-\mathbf{k}_{\mathrm{s}}\right)\cdot\mathbf{r}} {{d}}^3\mathbf{r} + \mathrm{h.c.}\right),
\end{align}
where $\delta\omega=2\omega_{\mathrm{p}}-\omega_\mathrm{pr}-\omega_\mathrm{s}$.
This Hamiltonian is equivalent to what is obtained in both PDC \cite{walborn2010spatial} and FWM \cite{nirala2023information}.
There are many routes that one may follow from this point.
First, we proceed in a manner that highlights the role of the spatial structure of the light fields in the position domain. For simplicity, we define the re-scaled operators $\hat{a}^\dagger_{\mathbf{k}_j}\equiv\sqrt{\omega_{\mathbf{k}_j}}\hat{\mathfrak{a}}^\dagger_{\mathbf{k}_j}$, to write
\begin{align} \label{eq:H2}
    \hat{H}_I &= -\frac{\hbar(\mathcal{E}^0_{\mathrm{p}})^2}{4V} \sum_{\mathbf{k}_{\mathrm{pr}},\mathbf{k}_{\mathrm{s}}}\left(\vphantom{\int_\mathbb{V}}\hat{a}^\dagger_{\mathbf{k}_{\mathrm{pr}}}\hat{a}^\dagger_{\mathbf{k}_{\mathrm{s}}}e^{i\delta\omega t}\right. 
    \\
    &\left.\times\int{{d}}z{{d}}^2\mathbf{r}_\perp\chi^{(3)} \mathcal{V}^2_\mathrm{p}(\mathbf{r}) e^{-i\Delta kz}e^{i(\boldsymbol{\rho}_{\mathrm{pr}}+\boldsymbol{\rho}_{\mathrm{s}})\cdot\mathbf{r}_\perp}+\mathrm{h.c.}\right),\nonumber
\end{align}
where
\begin{align} \label{eq:long_PM}
    \Delta k=2k_{\mathrm{p}}-(k_\mathrm{pr}+k_\mathrm{s})\cos\theta
\end{align}
is the longitudinal phase-mismatch.

In order to highlight the multi-spatial-mode character of the four-wave mixing process, we would like to switch from the continuous space of transverse momentum variables $(\boldsymbol{\rho}_{\mathrm{pr}},\boldsymbol{\rho}_{\mathrm{s}})$ to the discrete space of paraxial modes.
One may choose any paraxial basis. Here, for convenience, we work with the Laguerre-Gaussian modes, due to their cylindrical symmetry, and the possibility to study cases involving optical orbital angular momentum and other particular constructions, such as OAM Poincaré spheres \cite{da2024poincare}.
To this end, we use the closure relation between LG modes, $u_{\ell,p}(\mathbf{r})$, and their angular spectra, $\tilde{u}_{\ell,p}(\boldsymbol{\rho},z)$, \cite{calvo2006quantum} [see Eq. (\ref{eq:Fourier_completeness}) from Appendix \ref{Quantization_paraxial}]:
\begin{align}
    e^{i\boldsymbol{\rho}\cdot\mathbf{r}_\perp} &=2\pi\sum_{\ell,p}\tilde{u}_{\ell,p}(\boldsymbol{\rho},z) u^*_{\ell,p}(\mathbf{r}_\perp,z),
\end{align}
to rewrite the exponential $e^{i(\boldsymbol{\rho}_{\mathrm{pr}}+\boldsymbol{\rho}_{\mathrm{s}})\cdot\mathbf{r}_\perp}$.
In what follows, we shall assume that the paraxial condition ($|\boldsymbol{\rho}_j|^2/2k^2_j\ll1$) is valid.
Taking the discrete summations on the wave-vectors into integrals, as
\begin{equation}
    \sum_{\mathbf{k}} \rightarrow \dfrac{V}{(2\pi)^3}\int{{d}}^3\mathbf{k},
\end{equation}
and restricting ourselves to the thin-medium approximation, such that the longitudinal variation of the pump field can be neglected, we arrive at
\begin{align} \label{eq:NLH}
    \hat{H}_{I} &= A\int{{d}}k_\mathrm{s}{{d}}k_\mathrm{pr} \,\mathrm{sinc}\left(\frac{\Delta k L}{2}\right)e^{i\delta\omega t} 
    \\
    &\times\sum_{\ell_{\mathrm{pr}},p_{\mathrm{pr}}}\sum_{\ell_{\mathrm{s}},p_{\mathrm{s}}}C^{\ell_{\mathrm{pr}},\ell_{\mathrm{s}}}_{p_{\mathrm{pr}},p_{\mathrm{s}}}(\omega)\,\hat{a}^\dagger_{\ell_{\mathrm{pr}},p_{\mathrm{pr}}}(k_\mathrm{pr})\,\hat{a}^\dagger_{\ell_{\mathrm{s}},p_{\mathrm{s}}}(k_\mathrm{s}) + \mathrm{h.c.} \nonumber,
\end{align}
where $A=-2\pi^3\hbar LV(\mathcal{E}^0_{\mathrm{p}})^2$, $\omega$ represents $\{\omega_\mathrm{pr},\omega_\mathrm{s}\}$, the creation operator of a photon in the mode $(\ell,p)$ with longitudinal wave-number $k_j$ is
\begin{align}
\label{eq:alp_rho1}
    \hat{a}^\dagger_{\ell,p}(k_j)=\frac{1}{(2\pi)^2}\int\hat{a}^\dagger(\boldsymbol{\rho},k_j)\tilde{u}_{\ell,p}(\boldsymbol{\rho}){{d}}^2\boldsymbol{\rho},
\end{align}
satisfying $[\hat{a}_{\ell,p}(k),\hat{a}^\dagger_{\ell^\prime,p^\prime}(k^\prime)]=\delta_{\ell,\ell^\prime}\delta_{p,p^\prime}\delta(k-k^\prime)$, and finally,
\begin{align}\label{eq:OLI_Q}
    C^{\ell_{\mathrm{pr}},\ell_{\mathrm{s}}}_{p_{\mathrm{pr}},p_{\mathrm{s}}}(\omega) &= \int \chi^{(3)}(\omega)\mathcal{V}_{\mathrm{p}}^2{u}^*_{\ell_{\mathrm{pr}},p_{\mathrm{pr}}}{u}^*_{\ell_{\mathrm{s}},p_{\mathrm{s}}} \big|_{z=0}{{d}}^2\mathbf{r}_\perp,
\end{align}
is the transverse overlap integral.
Equation (\ref{eq:OLI_Q}) can be intuitively interpreted. The probability of generating a given pair of modes on the signal and probe channels is determined by the overlap of these modes with the interaction medium and pump field distributions.
This is another standard result in FWM \cite{walker2012trans,offer2018spiral}, which is central in the classical regime \cite{da2022spatial,da2024poincare}, where it determines the selection rules that dictate the transverse dynamics of optical modes.

For an arbitrary pump mode, given in the general case by Eq. (\ref{eq:Vpump}), we may rewrite Eq. (\ref{eq:OLI_Q}) as a summation of the form
\begin{align}
    C^{\ell_{\mathrm{pr}},\ell_{\mathrm{s}}}_{p_{\mathrm{pr}},p_{\mathrm{s}}}(\omega) &=\sum_{l,q}\sum_{l^\prime,q^\prime}c_{l,q}c_{l^\prime,q^\prime} C^{l,l^\prime,\ell_{\mathrm{pr}},\ell_{\mathrm{s}}}_{q,q^\prime,p_{\mathrm{pr}},p_{\mathrm{s}}}(\omega)
\end{align}
where the amplitudes with additional mode indices is
\begin{align} \label{eq:C_T}
    C^{l,l^\prime,\ell_{\mathrm{pr}},\ell_{\mathrm{s}}}_{q,q^\prime,p_{\mathrm{pr}},p_{\mathrm{s}}}(\omega)=\int \chi^{(3)}(\omega){u}_{l,q}{u}_{l^\prime,q^\prime}{u}^*_{\ell_{\mathrm{pr}},p_{\mathrm{pr}}}{u}^*_{\ell_{\mathrm{s}},p_{\mathrm{s}}}{{d}}^2\mathbf{r}_\perp.
\end{align}
We then see that the spatial overlap integral of four paraxial modes -- in this case, Laguerre-Gaussian modes -- also appears in the quantum-mechanical treatment of FWM. Here, in addition to containing the selection rules that dictate the allowed modes in the generated fields, it plays an important role in quantifying the squeezing and entanglement properties of light fields \cite{lanning2018quantized,offer2018spiral}.
It is important to emphasize, as already observed in Ref. \cite{nirala2023information}, that the interaction Hamiltonian given by Eq. (\ref{eq:NLH}) is very similar to that of parametric down conversion \cite{walborn2010spatial,baghdasaryan2022generalized}, except that for the FWM process, the convolution involves the angular spectra of the two pumping photons instead of the angular spectrum of the single pumping photon as in the case of PDC.

\subsection{Influence of the interaction medium geometry}
\begin{figure}[t!]
    \centering
    \includegraphics[width=0.9\linewidth]{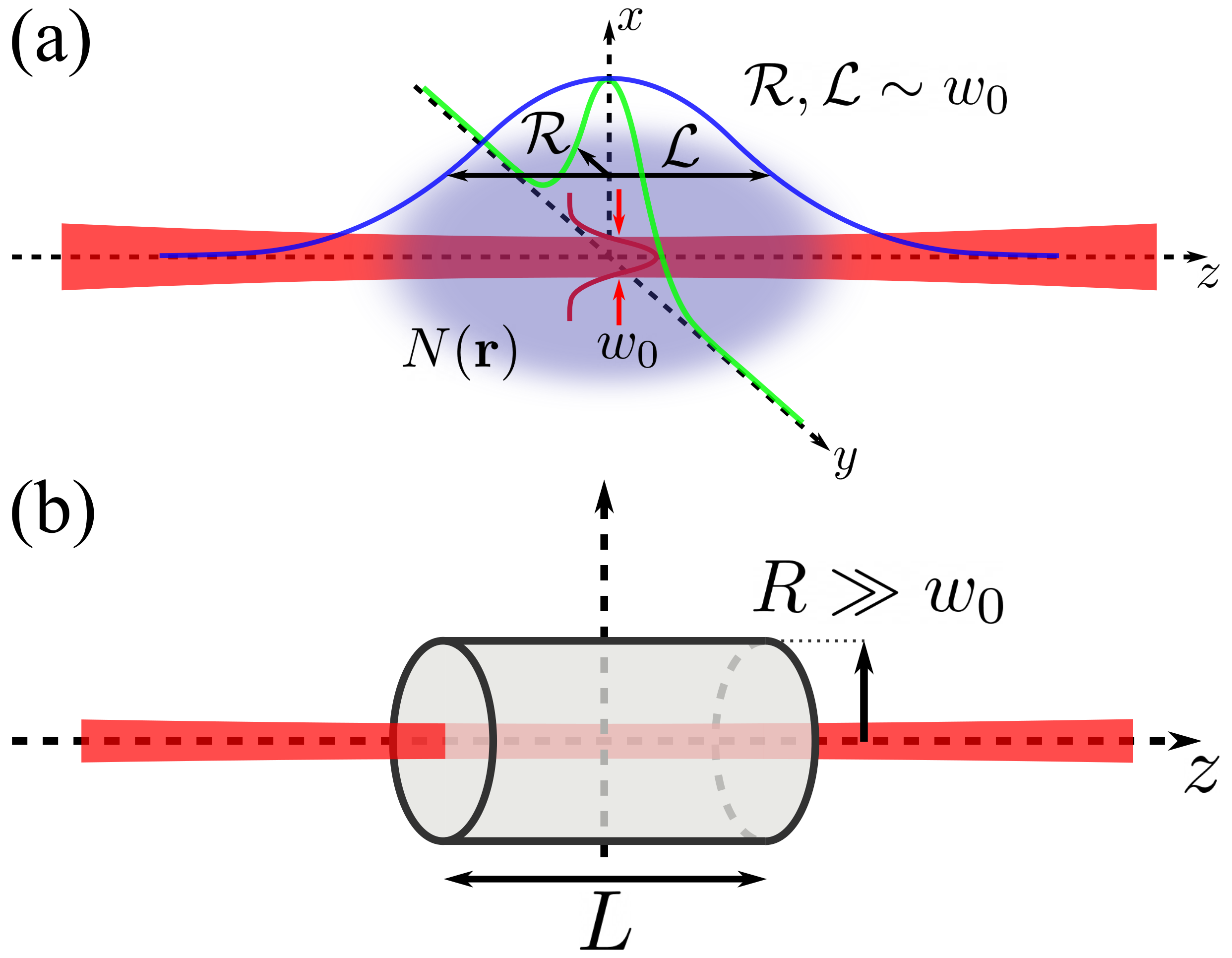}
    \caption{(a) Representation of the spatial distribution of atomic density in the case of a sample of cold atoms, $N_\mathrm{cloud}(\mathbf{r})$, as given by Eq. (\ref{eq:mu_cloud}). The cloud characteristic lengths $\mathcal{R}$ and $\mathcal{L}$ may be comparable to the optical waist $w_0$. (b) Depiction of an optical field incident on a glass cell containing a heated sample of atomic vapor, with a radius much larger than the beam waist parameter, $R\gg w_0$, in such a way that we may consider a uniform density of atoms along the cell length $L$.}
    \label{fig:atom_dens}
\end{figure}

The spatial dependence of the nonlinear susceptibility $\chi^{(3)}$ couples the spatial and spectral degrees of freedom.
This dependence may come from the pump field, as considered in Ref. \cite{da2022spatial}, and also from the shape of the sample.
The susceptibility can be assumed to have the form $\chi^{(3)}(\mathbf{r})\propto N(\mathbf{r})$, where $N(\mathbf{r})$ is the spatial distribution of the atomic density. In the case of a sample of cold atoms, we may use \cite{osorio2008spatial}
\begin{align} \label{eq:mu_cloud}
    N_\mathrm{cloud}(\mathbf{r})=\frac{N_0\pi^{-\frac{3}{2}}}{\mathcal{R}^2\mathcal{L}/2}\exp{\left(-\frac{r^2}{\mathcal{R}^2}-\frac{z^2}{(\mathcal{L}/2)^2}\right)},
\end{align}
where $N_0$ is the total number of atoms, and $\mathcal{R}$ and $\mathcal{L}$ are the transverse and longitudinal characteristic extensions of the atomic cloud.
In the case of a vapor cell, since the cell diameter is usually much larger than the transverse dimensions of the participating beams, we may neglect the transverse variation of the spread function inside the interaction region, and for simplicity, we consider
\begin{align}
    N_\mathrm{cell}(\mathbf{r})=\frac{N_0}{A_t L}\big[\Theta(z+L/2)-\Theta(z-L/2)\big],
\end{align}
where $L$ is the length of the vapor cell, $A_t$ is its cross-sectional area, and $\Theta(\cdot)$ is the Heaviside step function.
Figure \ref{fig:atom_dens} illustrates the two situations. In upcoming sections, we examine how a nonuniform distribution of atomic density affects the correlations between the generated fields.

For now, we shall consider a spatially uniform susceptibility over the interaction volume and in this manner $\chi^{(3)}$ factors out of the integral in Eq. (\ref{eq:C_T}). We can write:
\begin{align} \label{eq:C_A}
    C^{l,l^\prime,\ell_{\mathrm{pr}},\ell_{\mathrm{s}}}_{q,q^\prime,p_{\mathrm{pr}},p_{\mathrm{s}}}(\omega)=\chi^{(3)}(\omega)\Lambda^{l,l^\prime,\ell_{\mathrm{pr}},\ell_{\mathrm{s}}}_{q,q^\prime,p_{\mathrm{pr}},p_{\mathrm{s}}},
\end{align}
where
\begin{align}
    \Lambda^{l,l^\prime,\ell_{\mathrm{pr}},\ell_{\mathrm{s}}}_{q,q^\prime,p_{\mathrm{pr}},p_{\mathrm{s}}}=\int{u}_{l,q}{u}_{l^\prime,q^\prime}{u}^*_{\ell_{\mathrm{pr}},p_{\mathrm{pr}}}{u}^*_{\ell_{\mathrm{s}},p_{\mathrm{s}}}{{d}}^2\mathbf{r}_\perp,
\end{align}
is the transverse overlap of four LG modes, which can be readily calculated \cite{da2022spatial,da2024poincare}.

\section{The biphoton state in position representation}
\label{BPSPOS}

With the interaction Hamiltonian at hand, let us now move to the state of the quantum system at time $t$, $\ket{\psi(t)}$, which by solving Schrödinger's equation can be obtained as
\begin{align}
    \ket{\psi(t)}=\hat{U}(t)\ket{\psi(0)},
\end{align}
where $\hat{U}(t)$ is the time evolution operator
\begin{align}
    \hat{U}(t) &= \exp\left(-\frac{i}{\hbar}\int^t_0 {{d}}\tau \hat{H}_I(\tau)\right), \nonumber
    \\
    &\simeq \mathds{1} - \frac{i}{\hbar}\int^t_0 {{d}}\tau \hat{H}_I(\tau).
\end{align}
In what follows, we consider a quasi phase-matched configuration, $\Delta k\approx 0$, in such a way that the sinc function in the interaction Hamiltonian, $\hat{H}_I$ [Eq. (\ref{eq:NLH})], may be approximated by unity.
If we suppose that the interaction is turned on at time $t=0$, and that it lasts for a time much longer than any relevant time scale of the system, we can extend the integration limits to $\pm\infty$. This gives:
\begin{align}
    \int^\infty_{-\infty} {{d}}\tau e^{i(2\omega_{\mathrm{p}}-\omega_\mathrm{pr}-\omega_\mathrm{s})\tau}=2\pi\delta(2\omega_{\mathrm{p}}-\omega_\mathrm{pr}-\omega_\mathrm{s}).
\end{align}
Then, with the initial state as the vacuum, $\ket{0}$, the first-order approximation to the state at time $t$ can be written as
\begin{align}
    \ket{\psi(t)}&=\ket{0}+\mathcal{A}\int{{d}}\omega_\mathrm{pr}\sum_{\ell_{\mathrm{pr}},p_{\mathrm{pr}}}\sum_{\ell_{\mathrm{s}},p_{\mathrm{s}}}C^{\ell_{\mathrm{pr}},\ell_{\mathrm{s}}}_{p_{\mathrm{pr}},p_{\mathrm{s}}}(\omega_\mathrm{pr})
    \\
    &\quad\quad\quad\quad\quad\quad\quad\quad\quad\times\ket{\ell_{\mathrm{pr}},p_{\mathrm{pr}};\omega_\mathrm{pr}}\ket{\ell_{\mathrm{s}},p_{\mathrm{s}};\omega^\prime}, \nonumber
\end{align}
where $\mathcal{A}=-2\pi iA/\hbar$ is a constant and $\omega^\prime=2\omega_{\mathrm{p}}-\omega_{\mathrm{pr}}$.
The second term is the two-photon state, $\ket{\Psi}$.
With the substitution $\omega_\mathrm{pr} = \omega^0_\mathrm{pr} + {\Omega}$, where $\omega^0_\mathrm{pr}$ is regarded as a central frequency of the probe field, and ${\Omega}$ is a shift from this central frequency, we can write
\begin{align} \label{eq:biphoton_space}
    \ket{\psi(t)}&=\ket{0}+\mathcal{A}\int{{d}}{\Omega}\sum_{\ell_{\mathrm{pr}},p_{\mathrm{pr}}}\sum_{\ell_{\mathrm{s}},p_{\mathrm{s}}}C^{\ell_{\mathrm{pr}},\ell_{\mathrm{s}}}_{p_{\mathrm{pr}},p_{\mathrm{s}}}({\Omega})
    \\
    &\quad\quad\quad\quad\quad\quad\times\ket{\ell_{\mathrm{pr}},p_{\mathrm{pr}};\omega^0_{\mathrm{pr}}+{\Omega}}\ket{\ell_{\mathrm{s}},p_{\mathrm{s}};\omega^0_{\mathrm{s}}-{\Omega}}, \nonumber
\end{align}
where $\omega^0_{\mathrm{s}}=2\omega_{\mathrm{p}}-\omega^0_{\mathrm{pr}}$ is the central frequency of the signal field.
Finally, we consider that the nonlinear process is spectrally narrow around the central frequencies $(\omega^0_{\mathrm{pr}},\omega^0_{\mathrm{s}})$, or around ${\Omega}=0$, and we may write $\ket{\psi}=\ket{0}+\mathcal{A}^\prime \ket{\Psi}$ \footnote{The normalized light-state is $\ket{\overline{\psi}}=\frac{\ket{0}+\mathcal{A}^\prime\ket{\Psi}}{\sqrt{1+|\mathcal{A}^\prime|^2\braket{\Psi|\Psi}}}$.}, where $\mathcal{A}^\prime$ is a constant and $\ket{\Psi}$, the biphoton state, is given by:
\begin{align} \label{eq:two_photon_state}
    \ket{\Psi} = \sum_{\ell_{\mathrm{pr}},p_{\mathrm{pr}}}\sum_{\ell_{\mathrm{s}},p_{\mathrm{s}}}C^{\ell_{\mathrm{pr}},\ell_{\mathrm{s}}}_{p_{\mathrm{pr}},p_{\mathrm{s}}}\ket{\ell_{\mathrm{pr}},p_{\mathrm{pr}};\omega^0_{\mathrm{pr}}}_{\mathrm{pr}}\ket{\ell_{\mathrm{s}},p_{\mathrm{s}};\omega^0_{\mathrm{s}}}_{\mathrm{s}}.
\end{align}
This form, which describes an entangled bipartite quantum state, is a standard result in both PDC \cite{baghdasaryan2021justifying,baghdasaryan2022generalized} and FWM \cite{walker2012trans,offer2018spiral}. In Ref. \cite{offer2018spiral}, which investigates the quantum spiral bandwidth in a cascade FWM process, it is intuitively assumed that the generated light-state has the form given by Eq. (\ref{eq:two_photon_state}). Here, we outlined the steps needed to arrive at this conclusion.

\subsection{Entanglement properties of the biphoton state}
\label{entanglement_props}

In this section, we will outline some measures of entanglement commonly used in the study of biphotons. We will also detail how to compute these measures considering a particular subspace of spatial modes.
The distribution of coincidence amplitudes fully determines the biphoton state, as well as its entanglement and spatial correlation properties.
Thus, our calculations will require the set of coefficients $C^{\ell_{\mathrm{pr}},\ell_{\mathrm{s}}}_{p_{\mathrm{pr}},p_{\mathrm{s}}}$ in a specific subspace of modes.
An interesting quantity to analyze is the spiral bandwidth (SBW) of the two-photon state, defined as the standard deviation of the $\ell$--distribution \cite{torres2003quantum}. It is associated with the amount of entanglement between the fields $\mathrm{pr}$ and $\mathrm{s}$, and is directly related to the number of modes contributing to the entangled photon state. The SBW of the photon-pair generated in PDC has been extensively studied \cite{torres2003quantum,miatto2011full,yao2011angular}, and in FWM it was investigated in Ref. \cite{offer2018spiral}.

Following Ref. \cite{offer2018spiral}, it will be useful to define the biphoton $\ell$--distribution as a trace over the radial indices on the coincidence amplitudes:
\begin{align} \label{eq:PLL}
P_{\ell,\ell^\prime}\equiv\sum_{p,p^\prime}|C^{\ell,\ell^\prime}_{p,p^\prime}|^2.
\end{align}
For a pump given by a pure $u_{l,q}$ LG mode, due to OAM conservation, Eq. (\ref{eq:PLL}) becomes
\begin{align}
    P_{\ell,\ell_\mathrm{T}-\ell}=\sum_{p,p^\prime}|C^{l,l,\ell,\ell_\mathrm{T}-\ell}_{q,q,p,p^\prime}|^2,
\end{align}
where $\ell_\mathrm{T}=2l$ is the total OAM pumped into the system.
The SBW can then be expressed as:
\begin{align}
    \Delta\ell(\ell_\mathrm{T}) &= \sqrt{{\sum_{\ell}}\,\ell^2 P_{\ell,\ell_\mathrm{T}-\ell}-\left({\sum_{\ell}}\,\ell\, P_{\ell,\ell_\mathrm{T}-\ell}\right)^2}.
\end{align}
The associated entanglement entropy is \cite{leach2012secure,offer2018spiral}:
\begin{align} \label{eq:ent_entropy}
    {S}(\ell_\mathrm{T})=-\sum_{\ell}P_{\ell,\ell_\mathrm{T}-\ell}\log_2{(P_{\ell,\ell_\mathrm{T}-\ell})}.
\end{align}
We can explore the dependency of the SBW and the entanglement entropy with any system parameter, such as phase-mismatch, pumped OAM, medium length, beam waist, etc.

\begin{figure}[t!]
    \centering
    \includegraphics[width=0.9\linewidth]{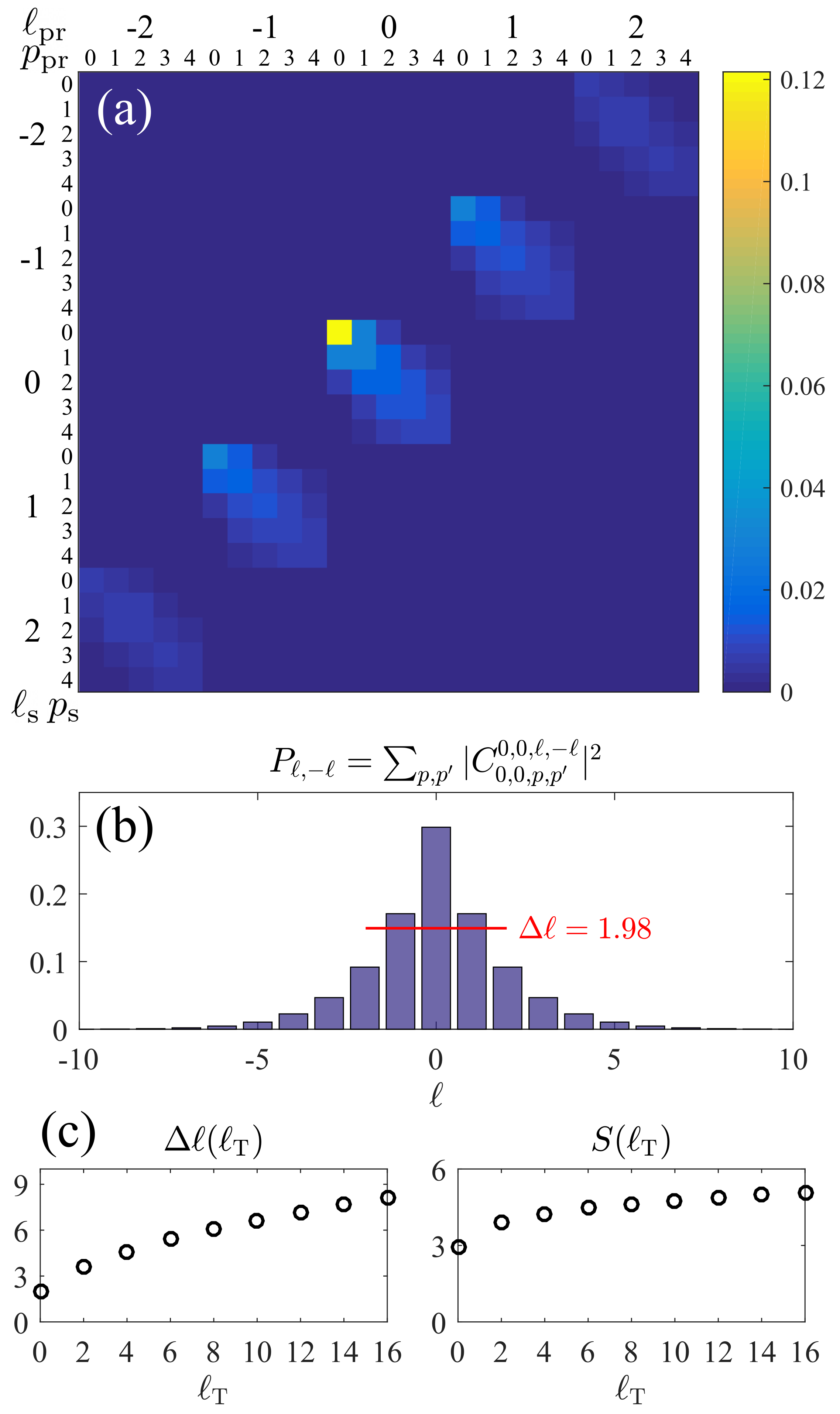}
    \caption{(a) Distribution of coincidence amplitudes, $|C^{\ell_{\mathrm{pr}},\ell_{\mathrm{s}}}_{p_{\mathrm{pr}},p_{\mathrm{s}}}|^2$, in the case of a Gaussian pump with a beam waist $w_0=1\,\mathrm{mm}$ exciting a thin nonlinear sample (the medium length does not significantly influence the calculations) of uniform transverse profile. (b) OAM distribution $P_{\ell,-\ell}$ of the biphoton state distribution shown in (a) evidencing a finite SBW $\Delta\ell\approx2$. (c) Variation of the SBW (left) and of the associated entanglement entropy (right) with the total OAM pumped into the system $\ell_\mathrm{T}$. The results were calculated using Eq. (\ref{eq:C_T}) considering the subspace of spatial modes $\mathbb{S}(2,4)$.}
    \label{fig:Gaussian_full}
\end{figure}

For the simplest case of a Gaussian pump, $\ell_\mathrm{T}=0$, we show in Fig. \ref{fig:Gaussian_full}(a) the amplitudes $C^{\ell_{\mathrm{pr}},\ell_{\mathrm{s}}}_{p_{\mathrm{pr}},p_{\mathrm{s}}}$ in the subspace $\mathbb{S}(l_{\mathrm{max}},p_{\mathrm{max}})$, spanned by the topological charges and radial indices $(\ell_\mathrm{pr},\ell_\mathrm{s})\in\{-l_\mathrm{max},...,l_\mathrm{max}\}$, $(p_\mathrm{pr},p_\mathrm{s})\in\{0,1,...,p_\mathrm{max}\}$, with $l_\mathrm{max}=2,p_\mathrm{max}=4$.
The total number of modes $\ket{\ell_\mathrm{pr},p_\mathrm{pr}}\ket{\ell_\mathrm{s},p_\mathrm{s}}$ constituting each subspace is equal to $M=[(2l_\mathrm{max}+1)\times(p_\mathrm{max}+1)]^2$.
It is interesting to note that even when the pump does not carry OAM, the light-state that emerges as a result of the nonlinear interaction is constituted by multiple nonzero OAM modes \cite{pan2019orbital}, leading to a nonzero SBW [see the $\ell$--distribution shown in Fig. \ref{fig:Gaussian_full}(b)].
Furthermore, for a non Gaussian pump, both the SBW and the entanglement entropy increase with the total pumped OAM, $\ell_\mathrm{T}$, as can be seen in Fig. \ref{fig:Gaussian_full}(c).
This is consistent with the results of Refs. \cite{offer2018spiral}, where a cascade FWM process generating blue light was considered.

\begin{figure}[h!]
    \centering
    \includegraphics[width=1\linewidth,trim={4cm 14.5cm 4cm 8cm}]{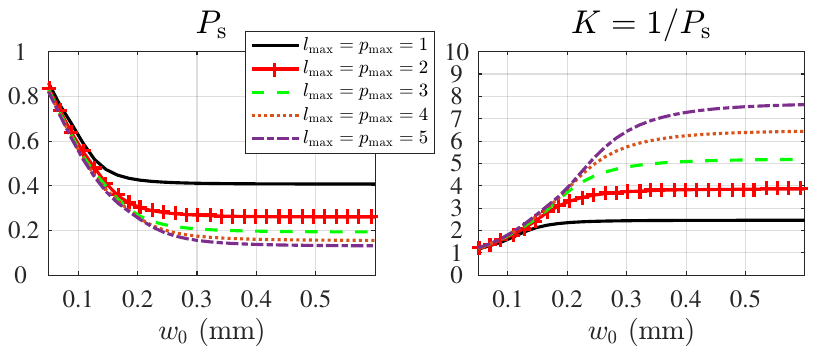}
    \caption{Purity of the partially traced biphoton state considering a Gaussian pump exciting a medium of length $L=5\,\mathrm{cm}$ restricted to subspaces $\mathbb{S}(l_\mathrm{max},p_\mathrm{max})$ of different sizes as a function of the waist $w_0$ (left), and the corresponding estimated Schmidt ranks in each subspace (right).}
    \label{fig:PK_subspace}
\end{figure}

We can also examine the spatial entanglement between the two generated photons by evaluating the purity of the reduced density matrix \cite{osorio2008spatiotemporal,baghdasaryan2022generalized}. Let $\hat{\rho}_\Psi = \ket{\Psi}\bra{\Psi}$ be the biphoton density operator, then the reduced state for the signal field is obtained by tracing over the probe field space $\hat{\rho}_{\Psi,\mathrm{s}}=\mathrm{tr}_\mathrm{pr}(\hat{\rho}_\Psi)$.
We are dealing only with spatial degrees of freedom, and therefore the trace is performed with respect to the discrete space of paraxial modes, i.e., $\mathrm{tr}(\hat{O})=\sum_{l,q}\bra{l,q}\hat{O}\ket{l,q}$.
Then, we obtain for the purity $P_\mathrm{s}$ \cite{schneeloch2016introduction,baghdasaryan2022generalized,osorio2008spatiotemporal}:
\begin{align} \label{eq:P_Psi}
    P_\mathrm{s}&=\mathrm{tr}(\hat{\rho}^2_{\Psi,\mathrm{s}}), \nonumber
    \\
    &=\sum C^{\ell,l}_{p,q}(C^{\ell^\prime,l}_{p^\prime,q})^*C^{\ell^\prime,l^\prime}_{p^\prime,q^\prime}(C^{\ell,l^\prime}_{p,q^\prime})^*,
\end{align}
where on the second line the summation is performed over all indices on the right-hand side.
The reduced state purity $P_\mathrm{s}$ is related to the Schmidt rank $K$ via:
\begin{align}\label{eq:Schmidt_P_Psi}
    K=\frac{1}{P_\mathrm{s}},
\end{align}
which is the number nonzero coefficients of the Schmidt decomposition of the quantum state \cite{law2004analysis,walborn2010spatial,schneeloch2016introduction}.
In this manner, we have another path to estimate the degree of spatial entanglement of the biphoton state and even examine the influence of parameters such as the total pumped OAM $\ell_\mathrm{T}$, the pump beam waist $w_0$, and the medium extension $L$.
We show in Fig. \ref{fig:PK_subspace} the purity $P_\mathrm{s}$ and the Schmidt rank $K=1/P_\mathrm{s}$ as a function of $w_0$ calculated using Eq. (\ref{eq:P_Psi}) and considering a Gaussian pump for different mode subspaces.
When we choose to project the Hamiltonian on the discrete basis of paraxial modes, the quantum state obtained is an approximation of the full state, i.e., the state given in momentum space with the appropriate biphoton amplitude function.
Consequently, all of the quantities we calculate on a given subset of modes become approximations of those we would get by considering the full state.
In fact, this procedure automatically accounts for an inherent limitation of real optical systems, as the size of the subset can be seen as a measure of the finite spatial bandwidth of the detection setup.

\subsection{Spatial correlation function of the biphoton state}
\label{SpatialCorrFunction}

\begin{figure*}[hbt!]
    \centering
    \includegraphics[width=1\linewidth,trim={0cm 0.75cm 0cm 0cm}]{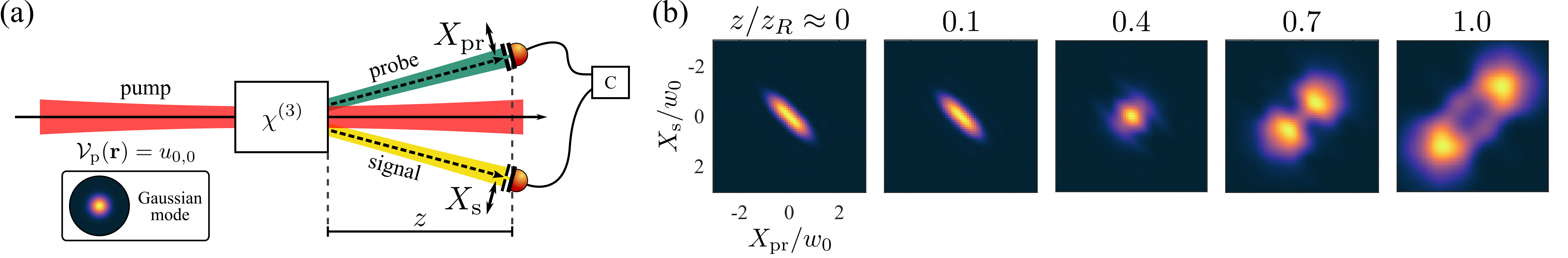}
    \caption{(a) Configuration for the detection of the spatial coincidence profiles in the case where we introduce spatial resolution by placing tight pinholes in front of detectors with a single translational degree of freedom. C corresponds to a coincidence counting system. (b) Evolution of the resulting coincidence profiles $g^{(2)}(X_\mathrm{pr},X_\mathrm{s})$ from the medium exit, $z=L/2$, up to $z=z_R$.
    }
    \label{fig:coincidence_profile_prop}
\end{figure*}

Other peculiar characteristics of the biphoton state are better visualized when we evaluate spatially resolved quantities \cite{monken1998transfer,boyer2008entangled}. As an example, the intensity correlation function, or coincidence count rate, between the fields $\mathrm{pr}$ and $\mathrm{s}$, can be found \cite{monken1998transfer}
\begin{align} \label{eq:G_Psi_rc_rs}
{G}(\mathbf{r}_{\mathrm{pr}},\mathbf{r}_{\mathrm{s}})&=\langle\psi|\hat{E}^\dagger_{\mathrm{pr}}(\mathbf{r}_{\mathrm{pr}})\hat{E}^\dagger_{\mathrm{s}}(\mathbf{r}_{\mathrm{s}})\hat{E}_{\mathrm{s}}(\mathbf{r}_{\mathrm{s}})\hat{E}_{\mathrm{pr}}(\mathbf{r}_{\mathrm{pr}})|\psi\rangle, \nonumber
\\
&=|\braket{0|\hat{E}_{\mathrm{s}}(\mathbf{r}_{\mathrm{s}})\hat{E}_{\mathrm{pr}}(\mathbf{r}_{\mathrm{pr}})|\Psi}|^2,\nonumber
\\
&\propto|\Psi(\mathbf{r}_{\mathrm{pr}},\mathbf{r}_{\mathrm{s}})|^2,
\end{align}
where $\Psi(\mathbf{r}_{\mathrm{pr}},\mathbf{r}_{\mathrm{s}})$ is the spatial mode function, defined as
\begin{align} \label{eq:Psi_rpr_rs}
     \Psi(\mathbf{r}_{\mathrm{pr}},\mathbf{r}_{\mathrm{s}}) &= \braket{\mathbf{r}_{\mathrm{pr}},\mathbf{r}_{\mathrm{s}}|\Psi}, 
     \\
     & =\sum_{\ell_{\mathrm{pr}},p_{\mathrm{pr}}}\sum_{\ell_{\mathrm{s}},p_{\mathrm{s}}}C^{\ell_{\mathrm{pr}},\ell_{\mathrm{s}}}_{p_{\mathrm{pr}},p_{\mathrm{s}}}{u}_{\ell_\mathrm{pr},p_{\mathrm{pr}}}(\mathbf{r}_{\mathrm{pr}}){u}_{\ell_{\mathrm{s}},p_{\mathrm{s}}}(\mathbf{r}_{\mathrm{s}}).\nonumber
\end{align}
Equation (\ref{eq:Psi_rpr_rs}) is a generally complex-valued function of four spatial coordinates.
For arbitrary aperture functions $\mathbb{A}_{\mathrm{pr}}$ and $\mathbb{A}_{\mathrm{s}}$, describing the shape of the spatial filters in front of the detectors of the fields $\mathrm{pr}$ and $\mathrm{s}$, we can write:
\begin{align} \label{eq:g2AA}
g^{(2)}=\int \mathbb{A}_{\mathrm{pr}}(\mathbf{r}_{\mathrm{pr}};\mathbf{R}_\mathrm{pr})\mathbb{A}_{\mathrm{s}}(\mathbf{r}_{\mathrm{s}};\mathbf{R}_\mathrm{s}){G}(\mathbf{r}_{\mathrm{pr}},\mathbf{r}_{\mathrm{s}}){{d}}^2\mathbf{r}_{\mathrm{pr}}{{d}}^2\mathbf{r}_{\mathrm{s}},
\end{align}
where $\mathbf{R}_{j}=(X_j,Y_j)$ represents the center position of the filters.
Consider now that we wish to detect only a circular region on the plane of detection of the probe and signal fields. For a really small aperture size $a_c\ll w_0$ centered at $\mathbf{R}_j=(X_j,0)$, $j=\mathrm{pr},\mathrm{s}$, representing a situation where we place a tight pinhole in front of the beam with a translational degree of freedom in the $x$--direction, we may write
\begin{align} \label{eq:Ai_def}
    \mathbb{A}_{j}(\mathbf{r}_{j};X_{j})&=\delta(x_{j}-X_{j})\delta(y_{j}), \quad j=\mathrm{pr},\mathrm{s}.
\end{align}
Substituting Eq. (\ref{eq:Ai_def}) for $\mathbb{A}_{\mathrm{pr},\mathrm{s}}$ into Eq. (\ref{eq:g2AA}), we obtain
\begin{align}
    g^{(2)}(X_{\mathrm{pr}},X_{\mathrm{s}})={G}(X_{\mathrm{pr}},0;X_{\mathrm{s}},0),
\end{align}
and we see that the spatial correlation function is directly determined by the amplitude of the biphoton wave-function, Eq. (\ref{eq:Psi_rpr_rs}), evaluated on the plane $y_\mathrm{pr}=y_\mathrm{s}=0$.
In these steps, it was assumed that both signal and probe are detected at the same longitudinal position $z\geq L/2$.
In Figure \ref{fig:coincidence_profile_prop} we show the coincidence count spatial profiles calculated using Eqs. (\ref{eq:C_A}), (\ref{eq:G_Psi_rc_rs}) and (\ref{eq:g2AA}) in the configuration of point-like detectors for different propagation distances starting from $z=L/2$, up to $z=z_R$, where $z_R=\frac{1}{2}k_\mathrm{p}w^2_0$ is the pump Rayleigh range.
The coincidence profile changes with the propagation distance outside the interaction medium.
At the medium exit (near-field), there is a positive correlation between the transverse positions of the outgoing photon pair. This occurs because the two photons are generated at the same transverse location, implying a strong position correlation at the source plane \cite{kumar2021einstein}.
As the fields propagate through free space, the coincidence profile gradually evolves, and at sufficiently large distances (far-field), it exhibits a negative correlation. This reflects the conservation of transverse linear momentum and the emergence of momentum anti-correlations in the far-field \cite{kumar2021einstein}.
This near-field to far-field transition of the biphoton spatial structure has recently been confirmed in the context of SPDC \cite{dehghan2024biphoton}.

\begin{figure}[hbt!]
    \centering
    \includegraphics[width=0.9\linewidth,trim={0cm 0.4cm 0cm 0cm}]{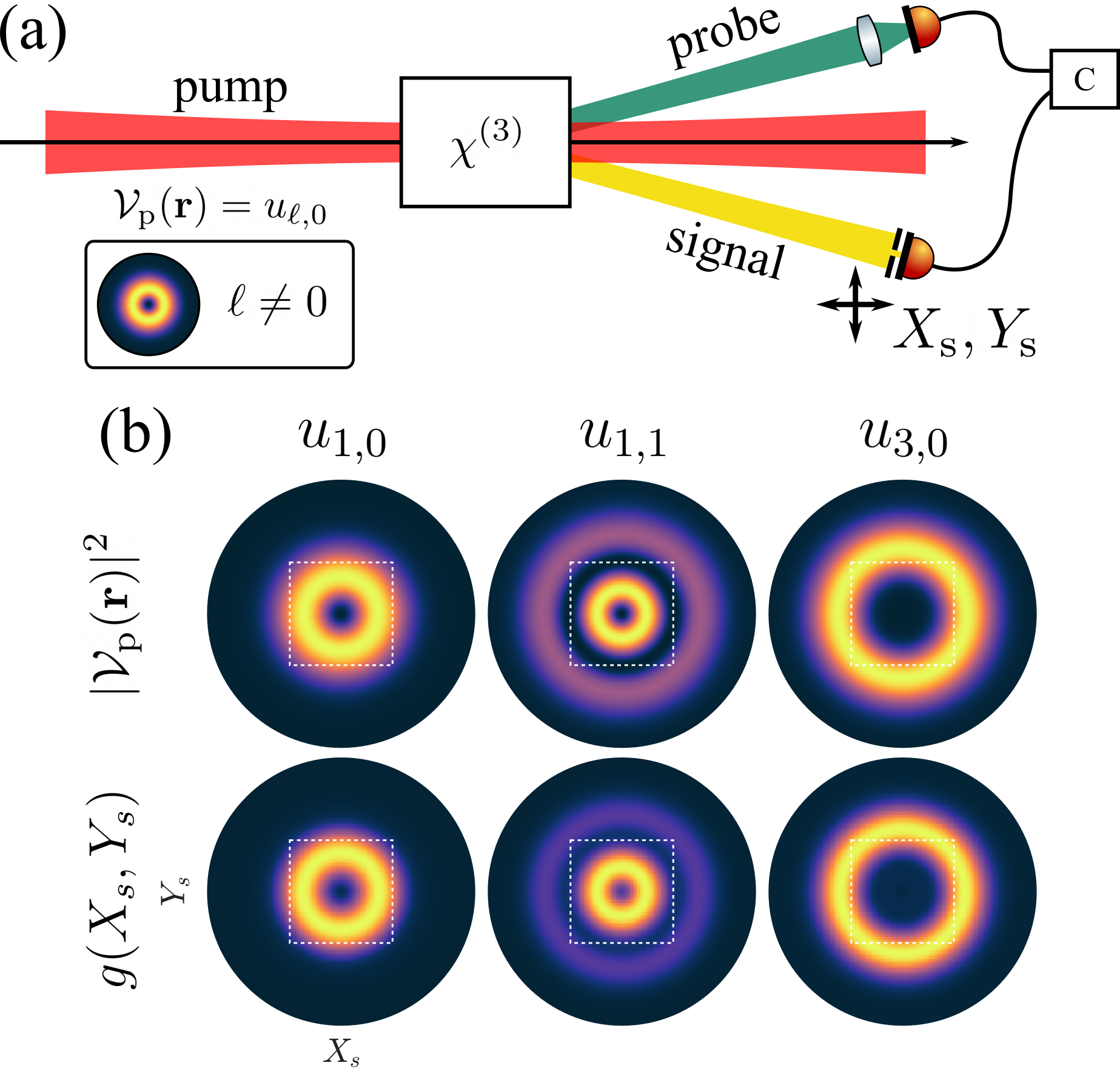}
    \caption{(a) Configuration for detecting the transfer of the pump spatial structure to the biphoton coincidence profile.
    The probe field is detected in its entirety, while the signal field is swept on the plane using a detector with two translational degrees of freedom.
    (b) Intensity profiles of the pump field carrying LG modes $u_{1,0},u_{1,1},u_{3,0}$ (top), and the corresponding coincidence profiles $g^{(2)}(X_\mathrm{s},Y_\mathrm{s})$ calculated using Eq. (\ref{eq:cr_det_total}) (bottom). For scale reference, the white boxes have sides equal to $w_0$.
    }
    \label{fig:pump_transfer}
\end{figure}

Suppose now that we detect the probe in its entirety ($\mathbb{A}_\mathrm{pr}=1$), and maintain a point-like detector on the plane of detection of the signal field centered at $\mathbf{R}_\mathrm{s}=(X_\mathrm{s},Y_\mathrm{s})$, now free to translate on the 2D plane. In this case, the signal aperture function is given by:
\begin{align}
    \mathbb{A}_\mathrm{s}(\mathbf{r}_\mathrm{s},\mathbf{R}_{\mathrm{s}})=\delta(\mathbf{r}_\mathrm{s}-\mathbf{R}_\mathrm{s}).
\end{align}
Substituting into Eq. (\ref{eq:g2AA}), the coincidence rate becomes
\begin{align} \label{eq:cr_det_total}
    &g^{(2)}(\mathbf{R}_{\mathrm{s}}) = \int \delta(\mathbf{r}_\mathrm{s}-\mathbf{R}_\mathrm{s}){G}(\mathbf{r}_{\mathrm{pr}},\mathbf{r}_{\mathrm{s}}){{d}}^2\mathbf{r}_{\mathrm{pr}}{{d}}^2\mathbf{r}_{\mathrm{s}},
    \\
    &\propto \sum_{\ell_{\mathrm{pr}},p_{\mathrm{pr}}}\sum_{\ell_{\mathrm{s}},p_{\mathrm{s}}}\sum_{\ell^\prime_{\mathrm{s}},p^\prime_{\mathrm{s}}} C^{\ell_{\mathrm{pr}},\ell_{\mathrm{s}}}_{p_{\mathrm{pr}},p_{\mathrm{s}}}(C^{\ell_{\mathrm{pr}},\ell^\prime_{\mathrm{s}}}_{p_{\mathrm{pr}},p^\prime_{\mathrm{s}}})^*{u}_{\ell_{\mathrm{s}},p_{\mathrm{s}}}(\mathbf{R}_{\mathrm{s}}){u}^*_{\ell^\prime_{\mathrm{s}},p^\prime_{\mathrm{s}}}(\mathbf{R}_{\mathrm{s}}). \nonumber
\end{align}
In Fig. \ref{fig:pump_transfer} we show the spatial coincidence count in this configuration, calculated using Eq. (\ref{eq:cr_det_total}) for different pump modes. These results evidence the transfer of the pump structure to the spatial coincidence profile $g^{(2)}$, a well-known result thoroughly explored in PDC \cite{monken1998transfer,walborn2010spatial}.
More recently, this behavior has also been evidenced in FWM using a camera able to capture images of both generated beams with a high repetition rate \cite{nirala2023information}.
It should be mentioned that, although not evident from the equation (\ref{eq:cr_det_total}), this behavior can be explicitly demonstrated. We will come back to this point in the following section, and provide a proof of this property.

\section{The biphoton state in momentum representation}
\label{biphoton_momentum}

Let us return to Eq. (\ref{eq:H2}), and outline a treatment in momentum space in order to capture other aspects and properties of the biphoton state.
As we move on, we will try to make connections with the results that we obtained in the previous Section.

Let us start by obtaining the biphoton state in the momentum representation.
With this objective, we define the pump angular spectrum $\widetilde{\mathrm{V}}(\boldsymbol{\rho},z)$ via $\widetilde{\mathrm{V}}(\boldsymbol{\rho},z)\equiv\mathcal{F}[\mathrm{V}(\mathbf{r})]=\frac{1}{2\pi}\int\mathrm{V}(\mathbf{r})e^{i\boldsymbol{\rho}\cdot\mathbf{r}_\perp}d^2\mathbf{r}_\perp$,
where $\mathrm{V}(\mathbf{r})=\mathcal{V}^2_\mathrm{p}(\mathbf{r})$ is the pump function, and $\mathcal{F}[\cdot]$ denotes the 2D spatial Fourier transform.
Substituting into Eq. (\ref{eq:H2}) and neglecting the slow longitudinal variation of $\widetilde{\mathrm{V}}$, it is straightforward to arrive at the expression \cite{baghdasaryan2021justifying,baghdasaryan2022generalized}:
\begin{align}
    &\hat{H}_I \propto \sum_{\mathbf{k}_{\mathrm{pr}},\mathbf{k}_{\mathrm{s}}}\hat{a}^\dagger_{\mathbf{k}_{\mathrm{pr}}}\hat{a}^\dagger_{\mathbf{k}_{\mathrm{s}}}\widetilde{\mathrm{V}}(\boldsymbol{\rho}_{\mathrm{pr}}+\boldsymbol{\rho}_{\mathrm{s}})\,\mathrm{sinc}\left(\frac{\Delta k L}{2}\right)e^{i\delta\omega t}+\mathrm{h.c.}, \nonumber
    \\
    &\rightarrow \int{{d}}^3\mathbf{k}_{\mathrm{pr}}{{d}}^3\mathbf{k}_{\mathrm{s}}\hat{a}^\dagger(\mathbf{k}_{\mathrm{pr}})\hat{a}^\dagger(\mathbf{k}_{\mathrm{s}})\Phi(\mathbf{k}_{\mathrm{pr}},\mathbf{k}_{\mathrm{s}})e^{i\delta\omega t}+\mathrm{h.c.}, \nonumber
\end{align}
where
\begin{align}
    \Phi(\mathbf{k}_{\mathrm{pr}},\mathbf{k}_{\mathrm{s}})=\widetilde{\mathrm{V}}(\boldsymbol{\rho}_{\mathrm{pr}}+\boldsymbol{\rho}_{\mathrm{s}})\mathrm{sinc}\left(\frac{\Delta k(k_\mathrm{pr},k_\mathrm{s}) L}{2}\right),
\end{align}
is a function that carries information from the pump distribution and also from phase-matching.
It should be noted that the structure of this function is the same as that which describes the light-state in PDC in second-order nonlinear crystals \cite{law2004analysis,walborn2010spatial,baghdasaryan2022generalized}.
It carries all of the relevant information for us to determine the twin beam state: the structure of the pump field and the phase-matching condition.

The quantum state at time $t$, in the first-order approximation, can be expressed as
\begin{align} \label{eq:BP_psi1}
    \ket{\tilde{\psi}(t)}=\ket{0}+\mathcal{B}\int{{d}}\omega_\mathrm{pr}&{{d}}^2\boldsymbol{\rho}_{\mathrm{pr}}{{d}}^2\boldsymbol{\rho}_{\mathrm{s}}\Phi\left(\boldsymbol{\rho}_{\mathrm{pr}},\frac{\omega_\mathrm{pr}}{c},\boldsymbol{\rho}_{\mathrm{s}},\frac{\omega^\prime}{c}\right) \nonumber
    \\
    &\times\ket{\boldsymbol{\rho}_{\mathrm{pr}},\omega_\mathrm{pr}}_{\mathrm{pr}}\ket{\boldsymbol{\rho}_{\mathrm{s}},\omega^\prime}_{\mathrm{s}},
\end{align}
where $\omega^\prime = 2\omega_{\mathrm{p}}-\omega_\mathrm{pr}$ and $\mathcal{B}$ is a constant.
Resorting once again to the assumption that the nonlinear process is narrow around the central frequencies $(\omega^0_\mathrm{pr},\omega^0_\mathrm{s})$, it is possible to represent the second term in Eq. (\ref{eq:BP_psi1}) as:
\begin{align}\label{eq:BP_psi2}
    \ket{\tilde{\Psi}} = \int{{d}}^2\boldsymbol{\rho}_{\mathrm{pr}}{{d}}^2\boldsymbol{\rho}_{\mathrm{s}}\Phi(\boldsymbol{\rho}_{\mathrm{pr}},\boldsymbol{\rho}_{\mathrm{s}})\ket{\boldsymbol{\rho}_{\mathrm{pr}},\omega_{\mathrm{pr}}}_{\mathrm{pr}}\ket{\boldsymbol{\rho}_{\mathrm{s}},\omega_{\mathrm{s}}}_{\mathrm{s}}.
\end{align}
This is the biphoton state on continuous momentum coordinates. To carry out our analysis in terms of spatial modes, we will now switch to the discrete space of mode indices.
Recognizing that $\ket{\boldsymbol{\rho}_j,\omega_j}_j=\hat{a}^\dagger(\boldsymbol{\rho}_j,k_j)\ket{0}$, we can use the inverse of relation (\ref{eq:alp_rho1}) in its quantum version,
\begin{align} \label{eq:a_LG}
    \hat{a}^\dagger(\boldsymbol{\rho},k) = (2\pi)^2\sum_{\ell,p}\hat{a}^\dagger_{\ell,p}(k)\tilde{u}^*_{\ell,p}(\boldsymbol{\rho}),
\end{align}
to project on the discrete space of paraxial LG modes.
Substituting Eq. (\ref{eq:a_LG}) into Eq. (\ref{eq:BP_psi1}), carrying a few simple steps, we obtain:
\begin{align} \label{eq:Psi_tilde}
    \ket{\tilde{\Psi}} = \sum_{\ell_{\mathrm{pr}},p_{\mathrm{pr}}}\sum_{\ell_{\mathrm{s}},p_{\mathrm{s}}}\widetilde{C}^{\ell_{\mathrm{pr}},\ell_{\mathrm{s}}}_{p_{\mathrm{pr}},p_{\mathrm{s}}}\ket{\ell_{\mathrm{pr}},p_{\mathrm{pr}};\omega^0_{\mathrm{pr}}}_{\mathrm{pr}}\ket{\ell_{\mathrm{s}},p_{\mathrm{s}};\omega^0_{\mathrm{s}}}_{\mathrm{s}},
\end{align}
where now the coincidence amplitude is written as
\begin{align} \label{eq:Ct_coeff}
    \widetilde{C}^{\ell_{\mathrm{pr}},\ell_{\mathrm{s}}}_{p_{\mathrm{pr}},p_{\mathrm{s}}}=\int{{d}}^2\boldsymbol{\rho}_{\mathrm{pr}}{{d}}^2\boldsymbol{\rho}_{\mathrm{s}}\Phi(\boldsymbol{\rho}_{\mathrm{pr}},\boldsymbol{\rho}_{\mathrm{s}})\tilde{u}^*_{\ell_{\mathrm{pr}},p_{\mathrm{pr}}}(\boldsymbol{\rho}_{\mathrm{pr}})\tilde{u}^*_{\ell_{\mathrm{s}},p_{\mathrm{s}}}(\boldsymbol{\rho}_{\mathrm{s}}).
\end{align}
For simplicity, from this point we omit any dependence of the biphoton function on the longitudinal wave-vector components.
This integral is very reminiscent of what we are used to encountering in position representation but with this approach, in the momentum representation, we automatically take into account the influence of the phase-mismatch and the medium length, as they are encoded in the sinc factor.
We will now focus once again on the specific case of a Gaussian pump beam, which allows us to perform calculations and discuss concrete results.

\subsection{Gaussian pump}

Let us now focus on the calculation of the coincidence amplitudes, as mentioned above, in the specific case of a Gaussian pump.
The Fourier transform of the squared Gaussian distribution can be obtained as $\mathcal{F}[{u}^2_{0,0}] = \frac{1}{2\pi}e^{-\rho^2w_0^2/8}$, and thus the pump angular spectrum is
\begin{align} \label{eq:V_mom}
    \widetilde{\mathrm{V}}(\boldsymbol{\rho}_{\mathrm{pr}}+\boldsymbol{\rho}_{\mathrm{s}}) = \frac{1}{2\pi}\exp\left(-\frac{w^2_0}{8}|\boldsymbol{\rho}_{\mathrm{pr}}+\boldsymbol{\rho}_{\mathrm{s}}|^2\right).
\end{align}
With the small-angle approximation, the following relation is valid \cite{schneeloch2016introduction}
\begin{align}
    k_\mathrm{pr}\sin\theta=\frac{|\boldsymbol{\rho}_{\mathrm{pr}}-\boldsymbol{\rho}_{\mathrm{s}}|}{2},
\end{align}
and along with Eq. (\ref{eq:long_PM}), we may write the longitudinal phase mismatch $\Delta k$ as
\begin{align}
    \Delta k = \frac{k_{\mathrm{p}}}{4k^2_{\mathrm{pr}}}|\boldsymbol{\rho}_{\mathrm{pr}}-\boldsymbol{\rho}_{\mathrm{s}}|^2.
\end{align}
Thus, in the degenerate setting, we have:
\begin{align} \label{eq:Phi_q1q2}
    \Phi(\boldsymbol{\rho}_{\mathrm{pr}},\boldsymbol{\rho}_{\mathrm{s}}) = \frac{1}{2\pi} e^{-\frac{w^2_0}{8}|\boldsymbol{\rho}_{\mathrm{pr}}+\boldsymbol{\rho}_{\mathrm{s}}|^2}\mathrm{sinc}\left(\frac{L}{8k}|\boldsymbol{\rho}_{\mathrm{pr}}-\boldsymbol{\rho}_{\mathrm{s}}|^2\right).
\end{align}
We note that by following the treatment in momentum space, the information regarding the thickness of the interaction medium is naturally included in the calculations.
In the upcoming sections, we will introduce modifications to account for the geometry of the sample in the particular case of a cloud of cold atoms.
The $\mathrm{sinc}(\cdot)$ function can be troublesome if used directly, and it can be written alternatively as
\begin{align} \label{eq:sinc_intz}
    \mathrm{sinc}\left(\frac{L}{8k}|\boldsymbol{\rho}_{\mathrm{pr}}-\boldsymbol{\rho}_{\mathrm{s}}|^2\right) = \frac{1}{L} \int^{L/2}_{-L/2}e^{-i\frac{z}{4k}|\boldsymbol{\rho}_{\mathrm{pr}}-\boldsymbol{\rho}_{\mathrm{s}}|^2}{{d}}z.
\end{align}
Then, using $|\boldsymbol{\rho}_{\mathrm{pr}}\pm\boldsymbol{\rho}_{\mathrm{s}}|^2 = \rho_\mathrm{pr}^2+\rho_\mathrm{s}^2\pm2\rho_\mathrm{pr}\rho_\mathrm{s}\cos(\varphi_\mathrm{pr}-\varphi_\mathrm{s})$, we can numerically calculate the coefficients of the biphoton state in the momentum representation given by Eq. (\ref{eq:Ct_coeff}):
\begin{align} \label{eq:T_qint}
    \widetilde{C}^{\ell_{\mathrm{pr}},\ell_{\mathrm{s}}}_{p_{\mathrm{pr}},p_{\mathrm{s}}}&\propto\int^{L/2}_{-L/2} dz\int {{d}}^2\boldsymbol{\rho}_{\mathrm{pr}}{{d}}^2\boldsymbol{\rho}_{\mathrm{s}}\,\tilde{u}^*_{\ell_{\mathrm{pr}},p_{\mathrm{pr}}}(\boldsymbol{\rho}_{\mathrm{pr}})\tilde{u}^*_{\ell_{\mathrm{s}},p_{\mathrm{s}}}(\boldsymbol{\rho}_{\mathrm{s}})\nonumber
    \\
    &\times e^{\left[-\alpha^2_+(z)(\rho^2_\mathrm{pr}+\rho^2_\mathrm{s})-2\rho_\mathrm{pr}\rho_\mathrm{s}\alpha^2_-(z)\cos(\varphi_\mathrm{pr}-\varphi_\mathrm{s})\right]},
\end{align}
where, $\alpha^2_\pm(z)\equiv(w^2_0/8\pm iz/4k)$.
With the set of coefficients calculated for a specific subset of mode indices, the state $\ket{\tilde{\Psi}}$, Eq. (\ref{eq:Psi_tilde}), is completely determined in the case of the Gaussian pump.

\begin{figure}[t!]
    \centering
    \includegraphics[width=0.95\linewidth,trim={4cm 8.7cm 5cm 8cm}]{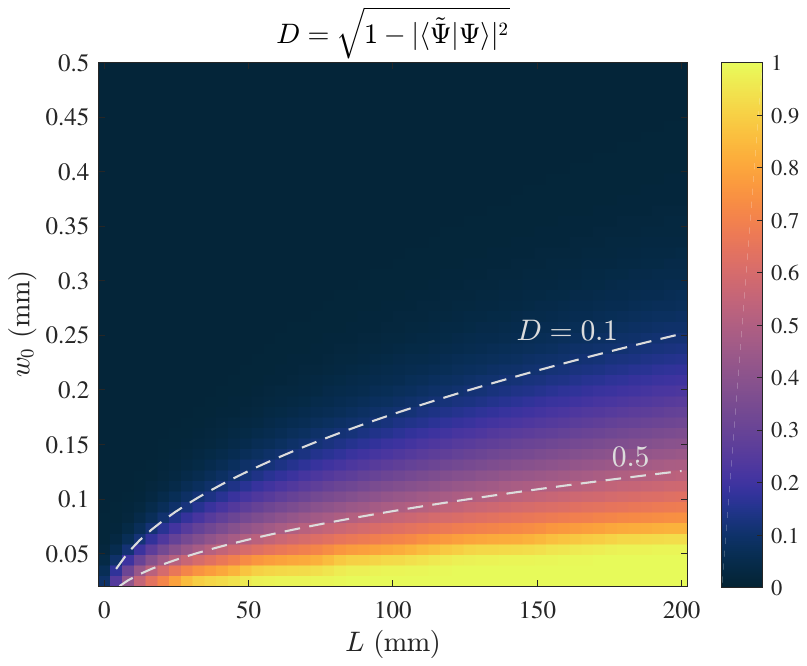}
    \caption{Trace distance ${{D}}=\sqrt{1-|\langle\tilde{\Psi}|\Psi\rangle|^2}$ as a function of $L$ and $w_0$ considering the subspace $\mathbb{S}(2,1)$ in the case of a Gaussian pump. The dashed lines represent the ${{D}}=0.1$ (upper) and ${{D}}=0.5$ (lower) contours, and these correspond to curves of the form $w_0\propto\sqrt{L}$.}
    \label{fig:D_wo_L}
\end{figure}

We can use a measure of distance between the position and momentum representations of the biphoton state via \cite{baghdasaryan2021justifying}
\begin{align} \label{eq:D_quantum}
    {{D}}(\ket{\tilde{\Psi}},\ket{\Psi}) &= \sqrt{1 - |\langle\tilde{\Psi}|\Psi\rangle|^2},
\end{align}
to probe the influence of the medium length $L$ and the beam waist $w_0$ on the biphoton state mode distribution.
In Fig. \ref{fig:D_wo_L} we show the dependence of the trace distance ${{D}}$ with the medium extension $L$ (considering the case of a vapor cell) and the beam waist $w_0$.
The form of the contours of constant $D$ coincides with the curves of constant ratio $L/z_R=2L/k_\mathrm{p}w^2_0$.
With this kind of result we can determine the configurations where the phase-matching function is relevant for the calculations.
In the limit of a thin medium, the contribution from the phase-matching term in the biphoton amplitude $\Phi$ becomes irrelevant.
Since in this limit we expect the states $\Ket{\Psi}$ and $\ket{\tilde{\Psi}}$ to coincide (${{D}}\rightarrow0$), the normalized overlap integrals in the position and momentum representations must be equal, $\widetilde{C}^{\ell_{\mathrm{pr}},\ell_{\mathrm{s}}}_{p_{\mathrm{pr}},p_{\mathrm{s}}}\overset{L/z_R\rightarrow0}{\longrightarrow} C^{\ell_{\mathrm{pr}},\ell_{\mathrm{s}}}_{p_{\mathrm{pr}},p_{\mathrm{s}}}$.
This is the case in configurations where the pair $(L,w_0)$ falls on the darker region of the map of Fig. \ref{fig:D_wo_L}.
For instance, in a medium with length of the order of a few centimeters, up to $L\sim 50\,\mathrm{mm}$ (typical size of Rb vapor cells), pumped by a laser tuned to the $5\mathrm{S}_{1/2}\rightarrow5\mathrm{P}_{3/2}$ transition of the $\mathrm{D}_2$ line of $\mathrm{Rb}^{87}$, our calculations suggest that for pump beam waists greater than $\sim 150$ $\upmu \mathrm{m}$, there is no sensible distinction between the states $\ket{\Psi}$ and $\ket{\tilde{\Psi}}$.
This behavior is analogous to what is observed in spontaneous parametric down-conversion under the thin-crystal approximation, and a similar conclusion is reached in Ref. \cite{baghdasaryan2021justifying}.
One difference we point out, that is of special interest for experimental purposes, is that in the FWM setting considered here, there seem to be more configurations $(w_0,L)$ that give smaller values for the trace distance $D$ (and therefore a greater fulfillment of the thin-medium condition). Making a direct comparison with the analogous figure of Ref. \cite{baghdasaryan2021justifying}, we see that the region of the map where $D<0.1$ is much larger here (if one considers the same scales for the axes).

If the interaction medium presents a nonuniform spatial distribution, as in Eq. (\ref{eq:mu_cloud}), the spread function $N(\mathbf{r})$ must be considered, in which case we modify the pump function to read $\widetilde{\mathrm{V}}(\boldsymbol{\rho},z)=\mathcal{F}\left[N(\mathbf{r})\mathcal{V}_{\mathrm{p}}^2(\mathbf{r})\right]$.
To account for this, we must make two changes to the calculations performed in this section. The first of them is to modify the width in Eq. (\ref{eq:V_mom}) as:
\begin{align}
    w_0\rightarrow w_\mathrm{eff}=\left(\frac{1}{w^2_0}+\frac{1}{2\mathcal{R}^2}\right)^{-\frac{1}{2}}=
    \frac{\sqrt{2}\xi w_0}{\sqrt{1+2\xi^2}},
\end{align}
where $\xi=\mathcal{R}/w_0$.
We can see $w_\mathrm{eff}$ as an effective beam waist, that takes into account the transverse overlap between the optical field and the atomic sample.
We observe that for $\mathcal{R}\rightarrow0$ (small cloud cross-section), we get $w_\mathrm{eff}\rightarrow0$, indicating a negligible light-atom interaction; and for $\mathcal{R}\rightarrow\infty$ (large cloud cross-section), $w_\mathrm{eff}\rightarrow w_0$, falling back to the case of a transversely uniform medium.
With this we can probe the influence of the dimensions of the atomic cloud on the mode structure of the generated photon-pair.
The second modification is to multiply the $z$ integrand in Eq. (\ref{eq:T_qint}) by the factor $e^{-4z^2/\mathcal{L}^2}$ to account for the longitudinal extension of the cloud $\mathcal{L}$.

We emphasize that the central difference in the calculations for FWM with respect to PDC lies in the fact that the pump contributes twice to the process. In this manner, we have two main changes: the width of the phase-matching function is modified; and the pump function is equal to the square of the pump mode, $\mathrm{V}(\mathbf{r}_\perp)= \mathcal{V}_{\mathrm{p}}^2(\mathbf{r}_\perp)$, effectively changing the width of the pump function both in the position and momentum spaces.
We can adapt the calculations from Refs. \cite{miatto2011full,baghdasaryan2022generalized} to obtain an (almost) analytical expression for $\widetilde{C}^{\ell_{\mathrm{pr}},\ell_{\mathrm{s}}}_{p_{\mathrm{pr}},p_{\mathrm{s}}}$, and even extend the treatment to accommodate an arbitrary pump structure. See Appendices \ref{arb_pump}-\ref{appendix_TT} for details on this procedure.

\subsection{Transfer of the pump structure to the photon-pair coincidence}

In Section \ref{SpatialCorrFunction} we calculated the spatial coincidence count using the biphoton function in the spatial representation, and saw in Fig. \ref{fig:pump_transfer} that the pump structure is transferred to the coincidence rate profile. We can now proof this result, starting with the biphoton state written in terms of continuous momentum variables, given by Eq. (\ref{eq:BP_psi2}).
It will be useful to write the biphoton function as $\Phi(\boldsymbol{\rho}_{\mathrm{pr}},\boldsymbol{\rho}_{\mathrm{s}})=\widetilde{\mathrm{V}}(\boldsymbol{\rho}_{\mathrm{pr}}+\boldsymbol{\rho}_{\mathrm{s}})\Delta(\boldsymbol{\rho}_{\mathrm{pr}}-\boldsymbol{\rho}_{\mathrm{s}})$, where $\Delta(\cdot)$ is the phase-matching function.
The coincidence count at zero time delay can be written as:
\begin{align} \label{eq:Psi_rz}
    {G}(\mathbf{r}_{\mathrm{pr}},z_{\mathrm{pr}};\mathbf{r}_{\mathrm{s}},z_{\mathrm{s}}) &= |\bra{0}\hat{E}_{\mathrm{s}}(\mathbf{r}_{\mathrm{s}},z_{\mathrm{s}})\hat{E}_{\mathrm{pr}}(\mathbf{r}_{\mathrm{pr}},z_{\mathrm{pr}})\ket{\tilde{\Psi}}|^2, \nonumber
    \\
    &\propto |\tilde{\Psi}(\mathbf{r}_{\mathrm{pr}},z_{\mathrm{pr}};\mathbf{r}_{\mathrm{s}},z_{\mathrm{s}})|^2,
\end{align}
where the spatial mode function $\tilde{\Psi}(\mathbf{r}_{\mathrm{pr}},z_{\mathrm{pr}};\mathbf{r}_{\mathrm{s}},z_{\mathrm{s}})$, after some manipulations, can be found as:
\begin{align} \label{eq:Z_rbrs}
    \tilde{\Psi}(\mathbf{r}_{\mathrm{pr}},&
    z_{\mathrm{pr}};\mathbf{r}_{\mathrm{s}},z_{\mathrm{s}})\propto\int{{d}}^2\boldsymbol{\rho}_{\mathrm{pr}}{{d}}^2\boldsymbol{\rho}_{\mathrm{s}} \widetilde{\mathrm{V}}(\boldsymbol{\rho}_{\mathrm{pr}}+\boldsymbol{\rho}_{\mathrm{s}})\Delta(\boldsymbol{\rho}_{\mathrm{pr}}-\boldsymbol{\rho}_{\mathrm{s}}) \nonumber
    \\
    &\quad\times e^{-i(\boldsymbol{\rho}_{\mathrm{pr}}\cdot\mathbf{r}_{\mathrm{pr}}+\boldsymbol{\rho}_{\mathrm{s}}\cdot\mathbf{r}_{\mathrm{s}})}e^{\left(i\frac{z_{\mathrm{pr}}}{2k_{\mathrm{pr}}}|\boldsymbol{\rho}_{\mathrm{pr}}|^2+i\frac{z_{\mathrm{s}}}{2k_{\mathrm{s}}}|\boldsymbol{\rho}_{\mathrm{s}}|^2\right)}.
\end{align}
Defining $\boldsymbol{\rho}_\pm=(\boldsymbol{\rho}_{\mathrm{pr}}\pm\boldsymbol{\rho}_{\mathrm{s}})/2$, and assuming $z_{\mathrm{pr}}=z_{\mathrm{s}}=z$ (detectors at the same longitudinal position), $k_{\mathrm{pr}}=k_{\mathrm{s}}=k$ (degenerate FWM process), allows us to arrive at the form:
\begin{align}
    \tilde{\Psi}(\mathbf{r}_{\mathrm{pr}},&z;\mathbf{r}_{\mathrm{s}},z)\propto \int \widetilde{\mathrm{V}}(\boldsymbol{\rho}_+)e^{i\frac{z}{4k}\rho^2_+}e^{-i\boldsymbol{\rho}_+\cdot\frac{(\mathbf{r}_{\mathrm{pr}}+\mathbf{r}_{\mathrm{s}})}{2}}{{d}}^2\boldsymbol{\rho}_+ \nonumber
    \\
    &\quad\quad\times\int \Delta(\boldsymbol{\rho}_-)e^{i\frac{z}{4k}\rho^2_-} e^{-i\boldsymbol{\rho}_-\cdot\frac{(\mathbf{r}_{\mathrm{pr}}-\mathbf{r}_{\mathrm{s}})}{2}}{{d}}^2\boldsymbol{\rho}_-.
\end{align}
Considering a quasi phase-matched regime, $\Delta k \approx 0$, such that $\Delta(\boldsymbol{\rho}_-)$ may be approximated by unity, at the medium exit, $z=0$, we obtain
\begin{align} \label{eq:Psi_V_Delta}
    \tilde{\Psi}(\mathbf{r}_{\mathrm{pr}},0;\mathbf{r}_{\mathrm{s}},0) \propto \mathrm{V}\left(\frac{\mathbf{r}_{\mathrm{pr}}+\mathbf{r}_{\mathrm{s}}}{2}\right)\delta\left(\frac{\mathbf{r}_{\mathrm{pr}}-\mathbf{r}_{\mathrm{s}}}{2}\right),
\end{align}
and by substituting Eq. (\ref{eq:Psi_V_Delta}) in Eq. (\ref{eq:Psi_rz}), we can write
\begin{align}
    {G}(\mathbf{r}_{\mathrm{pr}},0;\mathbf{r}_{\mathrm{s}},0) \propto \left|\mathrm{V}\left(\frac{\mathbf{r}_{\mathrm{pr}}+\mathbf{r}_{\mathrm{s}}}{2}\right)\right|^2\delta\left(\frac{\mathbf{r}_{\mathrm{pr}}-\mathbf{r}_{\mathrm{s}}}{2}\right).
\end{align}
In passing from Eq. (59) to Eq. (60) we assumed that the square of the Dirac delta function is proportional to the Dirac delta function itself, $(\delta(\mathbf{r}))^2\sim\delta(\mathbf{r})$.
Although this is not strictly valid in the mathematical sense, it is a common and physically motivated simplification, one that is often used, for instance, in calculations of quantum transition amplitudes \cite{zee2010quantum}.
Finally, the coincidence count profile obtained when the field $\mathrm{pr}$ is completely detected, Eq. (\ref{eq:cr_det_total}), can be expressed as:
\begin{align}
    g^{(2)}(\mathbf{R}_{\mathrm{s}}) &= \int \delta(\mathbf{r}_\mathrm{s}-\mathbf{R}_\mathrm{s}){G}(\mathbf{r}_{\mathrm{pr}},0;\mathbf{r}_{\mathrm{s}},0){{d}}^2\mathbf{r}_{\mathrm{pr}}{{d}}^2\mathbf{r}_{\mathrm{s}}, \nonumber
    \\
    &\propto |\mathrm{V}(\mathbf{R}_{\mathrm{s}})|^2,
\end{align}
clearly indicating that it is proportional to the pump function. We have therefore demonstrated that the pump structure is transferred to the coincidence count profile.
It is important to emphasize that, while we resorted to the description in momentum space to demonstrate this property of the biphoton state, the results shown in Fig. \ref{fig:pump_transfer} were calculated using the coincidence amplitudes in position space.

\subsection{Schmidt rank and entanglement}

In Ref. \cite{law2004analysis} the Schmidt rank of the (full) biphoton state generated in PDC was estimated by approximating the $\mathrm{sinc}(\cdot)$ phase-matching function as a Gaussian.
In doing so, the following analytical form was found \cite{law2004analysis}:
\begin{align} \label{eq:schmidt_law}
    K_G = \frac{1}{4}\Big(b\sigma_\perp+\frac{1}{b\sigma_\perp}\Big)^{2},
\end{align}
where $\sigma_\perp$ and $b^{-1}$ are the widths of the Gaussian pump function in wave-vector space,
and of the approximate Gaussian phase-matching function $\Delta_G(\boldsymbol{\rho})\propto e^{-b^2|\boldsymbol{\rho}|^2}$, respectively.

As we have seen throughout the last Sections, the calculations for the biphoton state generated in the FWM process are very similar with regard to PDC, with some differences arising mainly due to the fact that the pump and phase-matching functions are dictated by a squared contribution of the pump field.
These differences manifest precisely on the widths of the pump angular spectrum and phase-matching function.
For this reason, here we utilize Eq. (\ref{eq:schmidt_law}), taking into account the relevant modifications on the widths, to estimate the degree of entanglement for the FWM process.
We may encompass both cases by writing the widths $(\sigma_\perp,b)$ as:
\begin{align}
    \sigma^{(j)}_\perp &= \sqrt{4j}\frac{1}{w_0} = \sqrt{\frac{2jk}{z_R}}, \label{eq:sigma_K}
    \\
    b^{(j)} &= \gamma\sqrt{\frac{L}{4jk}}, \label{eq:b_K}
\end{align}
where $j=1$ for PDC, and $j=2$ for FWM, and the factor $\gamma\approx0.257$ ensures that the $\mathrm{sinc}(\cdot)$ function has the same width at half maximum as the Gaussian function considered in the approximation \cite{walborn2012generalized}.
Substituting Eqs. (\ref{eq:sigma_K}) and (\ref{eq:b_K}) in Eq. (\ref{eq:schmidt_law}), the Schmidt number for both processes can be expressed in terms of $\zeta=L/z_R$ as:
\begin{align} \label{eq:schmidt_ng}
    K_G (\zeta)&= \frac{\gamma^2}{8}\left(\sqrt{\zeta}+\frac{2}{\gamma^2}\frac{1}{\sqrt{\zeta}}\right)^2.
\end{align}
The ratio $L/z_R$, which appears in the classical picture as a parameter that allows us to establish different regimes of restriction on the FWM mode components through the Gouy phase-matching condition \cite{offer2021gouy}, is also connected with the entanglement properties of the photon pair in the quantum description.
Equation (\ref{eq:schmidt_ng}) predicts a large number of Schmidt modes (high entanglement) for $\zeta\ll1$, which goes to unity, $K_G\rightarrow1$ (low entanglement), with increasing $\zeta$, before rising again for very large $\zeta$.
This can be interpreted as follows. Larger medium extensions accommodate less momentum modes of light.
The uncertainty on the direction of the wave-vectors of the accepted modes is reduced, as compared with the situation in a thin-medium.

Considering $L=5\,\mathrm{cm}$ and varying the beam waist size $w_0$ from 50 $\upmu\mathrm{m}$ to $1\,\mathrm{mm}$, Eq. (\ref{eq:schmidt_ng}) predicts $K_G$ up to the order of hundreds of modes comprising the generated light-state.
The entanglement in this Gaussian approximation serves as a lower bound \cite{walborn2010spatial}, and therefore one would expect $K\geq K_G$ to hold true.
However, this estimate does not necessarily represent the actual degree of entanglement that is achievable in a realistic scenario.
In practice, the number of detectable entangled modes can be sensibly reduced, even by orders of magnitude \cite{walborn2010spatial,van2006effect,venkataraman2024unveiling}.
This is in accordance with the lower number of spatial modes estimated through our calculations using Eqs. (\ref{eq:P_Psi}) and (\ref{eq:Schmidt_P_Psi}).

\section{Conclusions}
\label{Conclusions}

In conclusion, we have detailed the quantum theory of four-wave mixing in atomic media, highlighting the parallels and the differences with regard to the well-established theory of parametric down-conversion in crystals. Starting from the nonlinear interaction Hamiltonian, we arrived at the biphoton state describing the entangled photon pair generated in the parametric process.
We outlined some well-understood results on spatial correlations and entanglement in FWM.

In particular, our calculations numerically and explicitly showed that the pump structure can be transferred to the spatial coincidence profile of the FWM biphoton state.
Although this kind of capability has been extensively studied in PDC, only more recently it was explored and demonstrated in FWM with a higher degree of control.

Another interesting result our work evidences is the behavior of the coincidence profile under free-space propagation of the generated photon-pair from the near- to the far-field regions.
At the medium exit, the coincidence profile indicates a positive spatial correlation, and as we move away to the far-field, we verify a negative correlation. This is in agreement with experimental results both in PDC and in FWM.

Lastly, regarding the entanglement measures of the light-state generated in FWM, we presented calculations of spiral bandwidth, entropy of entanglement and Schmidt rank considering restricted subspaces of spatial modes.
We discussed how the size of the subspace considered for the calculations ultimately affects the estimated degree of entanglement.
We argued that this route automatically accounts for the restriction imposed by the finite bandwidth of the detection setup over the exact Schmidt number in the Gaussian approximation.

Our results provide a theoretical basis for future work in nonlinear and quantum optics, explicitly accounting for structured pump beams and spatially resolved detection.

\begin{acknowledgments}
This work was supported by Coordena\c{c}\~{a}o de Aperfei\c{c}oamento de Pessoal de N\'{\i}vel Superior (CAPES - PROEX Grant 534/2018, No.
23038.003382/2018-39) and Funda\c{c}\~{a}o de Amparo \`{a} Pesquisa do Estado de S\~{a}o Paulo (FAPESP - Grant 2021/06535-0). M. R. L. da Motta acknowledges financial support from CAPES (Grant 88887.623521/2021-00). S. S. Vianna acknowledges financial support from CNPq (Grant 307722/2023-6).
\end{acknowledgments}

\appendix

\section{The LG mode and its angular spectrum}
\label{LG_modes}

The Laguerre-Gaussian mode is an exact solution to the paraxial wave equation (PWE) in cylindrical coordinates. It is characterized by the pair of integers $(\ell,p)$, known as the topological charge and the radial index, respectively, with $\ell\in\mathbb{Z}$, and $p=0,1,2,\dots$. We denote the LG mode as:
\begin{align}
    u_{\ell,p}(\mathbf{r}) &= \frac{\mathcal{N}_{\ell,p}}{w(z)}r^{|\ell|}_w L_{p}^{|\ell|}\left(r^2_w\right) e^{-r^2_w/2}e^{i\ell\phi} e^{-i\frac{kr^2}{2R(z)} + i\Psi_{\mathrm{G}} (z)},
\end{align}
where the position vector in cylindrical coordinates is $\mathbf{r}=(r,\phi,z)$, $r_w\equiv \sqrt{2}r/w(z)$ is the rescaled radial coordinate, $\mathcal{N}_{\ell,p}=\sqrt{2p!/\pi(p+|\ell|)!}$ is a normalization constant, $L_{p}^{|\ell|}(\cdot)$ is the associated Laguerre polynomial, $w(z)=w_0\sqrt{1+(z/z_{R})^{2}}$ is the beam waist, $z_R$ is the Rayleigh range, $R(z)=z\left[1+(z_{R}/z)^{2}\right]$ is the curvature radius, and $\Psi_{\mathrm{G}}(z)=(N_{\ell,p}+1)\tan^{-1}(z/z_R)$ is the Gouy phase shift, with the total mode order defined as $N_{\ell,p}=2p+|\ell|$.

The angular spectrum of the LG mode at $z=0$ is given by its spatial Fourier transform
\begin{align} \label{eq:FT_z0}
    \tilde{u}_{\ell,p}(\boldsymbol{\rho}) &= \frac{1}{2\pi}\int u_{\ell,p}(\mathbf{r}_\perp,0)e^{i\boldsymbol{\rho}\cdot\mathbf{r}_\perp}{{d}}^2\mathbf{r}_\perp, \nonumber
    \\
    &=\mathcal{N}_{\ell,p} \dfrac{w_0}{2}\rho^{|\ell|}_w L_p^{|\ell|}\left(\rho^2_w\right) e^{i\ell\varphi}e^{-\rho^2_w/2}e^{i\frac{\pi}{2}N_{\ell,p}},
\end{align}
where $\boldsymbol{\rho}=(\rho,\varphi)$ is the transverse wave-vector, and $\rho_w\equiv \rho w_0/\sqrt{2}$. The PWE in momentum space describes the evolution of the angular spectrum, as:
\begin{align}
    -i\vartheta^2_\mathbf{k}\tilde{u}_{\ell,p}+\frac{\partial \tilde{u}_{\ell,p}}{\partial z} = 0,
\end{align}
where
\begin{align}
    \vartheta_\mathbf{k}\equiv |\boldsymbol{\rho}|/\sqrt{2}k
\end{align}
is a parameter that determines the degree of paraxiality.
The solution to this equation gives the propagation of the angular spectrum:
\begin{align}
    \tilde{u}_{\ell,p}(\boldsymbol{\rho},z) = \tilde{u}_{\ell,p}(\boldsymbol{\rho},0)e^{ik\vartheta^2_\mathbf{k}z}.
\end{align}


\section{Quantization of a paraxial electromagnetic field}
\label{Quantization_paraxial}


To establish a quantum description of the FWM process that contemplates the spatial degrees of freedom of light, we must represent the paraxial light fields within a second-quantized framework.
Therefore, here we outline the quantization of a paraxial electromagnetic field.
For a more rigorous procedure, one may consult Refs. \cite{aiello2005exact,calvo2006quantum}.

We start with the vector potential in free-space expanded in terms of plane-waves:
\begin{align} \label{eq:A_pw}
    \mathbf{A} = \sum_{\sigma,\mathbf{k}} \dfrac{1}{\sqrt{2\varepsilon_0\omega_{\mathbf{k}}V}} \Big(\boldsymbol{\epsilon}_{\sigma,\mathbf{k}} \mathfrak{a}_{\sigma,\mathbf{k}}e^{-i(\mathbf{k}\cdot\mathbf{r}-\omega_\mathbf{k}t)}+\mathrm{c.c.}\Big),
\end{align}
where, in addition to what was defined in the main text, $\boldsymbol{\epsilon}_{\sigma,\mathbf{k}}$, with $\sigma=1,2$, denotes a particular polarization basis orthogonal to the wave-vector $\mathbf{k}$.
We write $\mathbf{k}$ as $\mathbf{k}=k\mathbf{e}_z+\boldsymbol{\rho}$
and, without loss of generality, $k\geq0$, representing waves traveling along the positive $z$ direction.
Also, we separate the vector potential into positive and negative frequency components, $\mathbf{A}(\mathbf{r},t)=\mathbf{A}^{(+)}(\mathbf{r},t)+\mathbf{A}^{(-)}(\mathbf{r},t)$, where:
\begin{align} \label{eq:E_qr}
    \mathbf{A}^{(+)} = \dfrac{1}{\sqrt{2\varepsilon_0V}}\sum_{\sigma,\mathbf{k}}\frac{1}{\sqrt{\omega_\mathbf{k}}}\boldsymbol{\epsilon}_{\sigma,\mathbf{k}}\mathfrak{a}_{\sigma,\mathbf{k}}e^{-i(kz+\boldsymbol{\rho}\cdot\mathbf{r}_\perp-\omega_\mathbf{k}t)},
\end{align}
and $\mathbf{A}^{(-)}=[\mathbf{A}^{(+)}]^*$.
Now, the transverse position-momentum exponential $e^{i\boldsymbol{\rho}\cdot\mathbf{r}_\perp}$ can be expressed in terms of the following closure relation between the LG mode and its angular spectrum:
\begin{align} \label{eq:Fourier_completeness}
    e^{i\boldsymbol{\rho}\cdot\mathbf{r}_\perp} &= \int e^{i\boldsymbol{\rho}\cdot\mathbf{r}^\prime_\perp} \delta^{(2)}(\mathbf{r}_\perp-\mathbf{r}^\prime_\perp){{d}}^2\mathbf{r}^\prime_\perp, \nonumber
    \\
    &=\sum_{\ell,p} \int e^{i\boldsymbol{\rho}\cdot\mathbf{r}^\prime_\perp} u_{\ell,p}(\mathbf{r}^\prime_\perp,z) u^*_{\ell,p}(\mathbf{r}_\perp,z){{d}}^2\mathbf{r}^\prime_\perp, \nonumber
    \\
    &=2\pi\sum_{\ell,p}\tilde{u}_{\ell,p}(\boldsymbol{\rho},z) u^*_{\ell,p}(\mathbf{r}_\perp,z),
\end{align}
where the completeness relation of the LG modes was used.
Considering a continuous spectrum of wave-vectors, the summation on $\mathbf{k}$ becomes an integral as:
\begin{equation}
    \sum_{\mathbf{k},k\geq0} \rightarrow \dfrac{V}{(2\pi)^3}\iint_{\mathbb{R}^2}{{d}}^2\boldsymbol{\rho}\int^{+\infty}_{0}{{d}}k,
\end{equation}
and therefore we can write:
\begin{align} \label{eq:E_field_cig}
    \mathbf{A}^{(+)}&(\mathbf{r},t) = \dfrac{\mathcal{A}}{(2\pi)^2} \sum_{\sigma,\ell,p}\iint_{\mathbb{R}^2}{{d}}^2 \boldsymbol{\rho}\int^{\infty}_{0} \frac{{{d}}k}{\sqrt{\omega_\mathbf{k}}}\boldsymbol{\epsilon}_{\sigma,\mathbf{k}} \mathfrak{a}_{\sigma}(\boldsymbol{\rho},k) \nonumber
    \\
    &\,\,\,\,\,\times\tilde{u}^*_{\ell,p}(\boldsymbol{\rho},0) u_{\ell,p}(\mathbf{r}_\perp,z;k)e^{-i[k(1+\vartheta^2_\mathbf{k})z-c|\mathbf{k}|t]},
\end{align}
where $\mathcal{A}=\sqrt{V/2\varepsilon_0}$.
In the paraxial limit, $\vartheta_\mathbf{k}\ll1$, and $\boldsymbol{\epsilon}_{\sigma,\mathbf{k}}\simeq\boldsymbol{\epsilon}_{\sigma}$ ($\boldsymbol{\epsilon}^{(*)}_\sigma\cdot\mathbf{e}_z=0$), $\omega_\mathbf{k}\simeq ck=\omega_k$.
We then define \cite{calvo2006quantum}:
\begin{equation} \label{eq:a_slpk}
    \mathfrak{a}_{\sigma,\ell,p}(k) \equiv \dfrac{1}{(2\pi)^2} \int  \mathfrak{a}_{\sigma}(\boldsymbol{\rho},k) \tilde{u}^*_{\ell,p}(\boldsymbol{\rho}){{d}}^2\boldsymbol{\rho},
\end{equation}
as the amplitude of a field component with polarization $\sigma$ and wave-number $k$ that is described by an LG mode with indices $\ell$ and $p$.
Then, defining the re-scaled amplitudes $a_{\sigma,\ell,p}(k)\equiv\mathfrak{a}_{\sigma,\ell,p}(k)/\sqrt{\omega_k}$ we can write:
\begin{align} \label{eq:E_ready}
    \mathbf{A}^{(+)} = \mathcal{A}\sum_{\sigma,\ell,p}\int{{d}}k\,a_{\sigma,\ell,p}(k)\boldsymbol{\epsilon}_{\sigma} u_{\ell,p}(\mathbf{r})e^{-i(kz-\omega_kt)},
\end{align}
The vector potential as given by Eq. (\ref{eq:E_ready}) is now suitable to become a quantum-mechanical operator.
To this end, we promote the complex field amplitudes $a^{(*)}$ to bosonic annihilation and creation operators:
\begin{align}
    a^{(*)}_{\sigma,\ell,p}(k)\rightarrow\sqrt{\hbar}\,\hat{a}^{(\dagger)}_{\sigma,\ell,p}(k),
\end{align}
which satisfy the commutation relations:
\begin{equation}
    [\hat{a}_{\sigma,\ell,p}(k),\hat{a}^\dagger_{\sigma^\prime,\ell^\prime,p^\prime}(k^\prime)]=\delta_{\sigma,\sigma^\prime}\delta_{\ell,\ell^\prime}\delta_{p,p^\prime}\delta(k-k^\prime).
\end{equation}
The vector potential becomes an operator as:
\begin{align}
    \mathbf{A}^{(+)}&\rightarrow\hat{\mathbf{A}},
    \\
    \mathbf{A}^{(-)}&\rightarrow\hat{\mathbf{A}}^\dagger.
\end{align}
The electric and magnetic fields also become quantum operators $(\mathbf{E},\mathbf{B})\rightarrow(\hat{\mathbf{E}},\hat{\mathbf{B}})$.
The photon number states we work with in the main text can then be represented as:
\begin{align}
    \ket{\{n_{\sigma,\ell,p}(k)\}}&\equiv\ket{n_{1},n_{2},...}, \nonumber
    \\
    &=\prod_{j}\frac{(\hat{a}^\dagger_{g_j})^{n_{j}}}{\sqrt{n_{j}!}}\ket{0,0,...},
\end{align}
where $g_j=\{\sigma,\ell,p;k\}_j$ labels a particular combination of polarization, paraxial mode indices, and wave-number value; and $n_{j}$ is the number of photons in the mode $g_j$.
The results are analogous if one considers another paraxial basis $\{u_{m,n}\}$, such as the Hermite-Gaussian modes.

\section{Calculation of the coincidence amplitude in momentum space}

\subsection{Arbitrary pump}
\label{arb_pump}

With the objective of adapting the calculations in Section \ref{biphoton_momentum} to allow us to consider arbitrary pump modes, we first observe that the product of two LG beams can be expressed as a superposition of LG modes as $(u_{l,q}u_{l^\prime,q^\prime})=\sum_{m,n}s^{m,l,l^\prime}_{n,q,q^\prime}u_{m,n}$ \cite{kotlyar2022product}, where $s^{m,l,l^\prime}_{n,q,q^\prime}=\int (u_{l,q}u_{l^\prime,q^\prime})u^*_{m,n}{{d}}^2\mathbf{r}_\perp$.
Thus for an arbitrary pump structure $\mathcal{V}_{\mathrm{p}} = \sum_{l,q}c_{l,q}u_{l,q}$, we have:
\begin{align}
    \mathrm{V}(\mathbf{r}_\perp) = \mathcal{V}_{\mathrm{p}}^2(\mathbf{r}_\perp) &= \sum_{l,q;l^\prime,q^\prime}c_{l,q}c_{l^\prime,q^\prime}\Big[{u}_{l,q}(\mathbf{r}_\perp){u}_{l^\prime,q^\prime}(\mathbf{r}_\perp)\Big], \nonumber
    \\
    &=\sum_{m,n}a_{m,n}{u}_{m,n}(\mathbf{r}_\perp),
\end{align}
where
\begin{align}
\label{eq:amns}
    a_{m,n} = \sum_{l,q}\sum_{l^\prime,q^\prime}c_{l,q}c_{l^\prime,q^\prime}s^{m,l,l^\prime}_{n,q,q^\prime}.
\end{align}
The pump function in Fourier space is thus given by the superposition of LG angular spectra:
\begin{align}
    \widetilde{\mathrm{V}}(\boldsymbol{\rho})=\sum_{m,n}a_{m,n}\tilde{u}_{m,n}(\boldsymbol{\rho}).
\end{align}
With the set of coefficients $\{a_{m,n}\}$ at hand, the coincidence amplitudes can be expressed as
\begin{align}
    \label{eq:CT_amn}
    \widetilde{C}^{\ell_{\mathrm{pr}},\ell_{\mathrm{s}}}_{p_{\mathrm{pr}},p_{\mathrm{s}}} &= \sum_{m,n}a_{m,n}\int{{d}}^2\boldsymbol{\rho}_{\mathrm{pr}}{{d}}^2\boldsymbol{\rho}_{\mathrm{s}} \, \Delta(\boldsymbol{\rho}_{\mathrm{pr}}-\boldsymbol{\rho}_{\mathrm{s}})\tilde{u}_{m,n}(\boldsymbol{\rho}_{\mathrm{pr}}+\boldsymbol{\rho}_{\mathrm{s}})\nonumber
    \\
    &\quad\quad\quad\quad\quad\times\tilde{u}^*_{\ell_{\mathrm{pr}},p_{\mathrm{pr}}}(\boldsymbol{\rho}_{\mathrm{pr}})\tilde{u}^*_{\ell_{\mathrm{s}},p_{\mathrm{s}}}(\boldsymbol{\rho}_{\mathrm{s}}),\nonumber
    \\
    &=\frac{1}{L} \sum_{m,n}a_{m,n} \int^{\frac{L}{2}}_{-\frac{L}{2}}T^{\ell_{\mathrm{pr}},\ell_{\mathrm{s}},m}_{p_{\mathrm{pr}},p_{\mathrm{s}},n}(z){{d}}z,
\end{align}
%
where we used Eq. (\ref{eq:sinc_intz}) for the phase-matching function and:
    \begin{align} \label{eq:T_momentum}
    T^{\ell_{\mathrm{pr}},\ell_{\mathrm{s}},m}_{p_{\mathrm{pr}},p_{\mathrm{s}},n}(z) = &\int {{d}}^2\boldsymbol{\rho}{{d}}^2\boldsymbol{\rho}^\prime\exp{\left(-i\frac{z}{4k}|\boldsymbol{\rho}-\boldsymbol{\rho}^\prime|^2\right)} \nonumber
    \\
    &\times\tilde{u}_{m,n}(\boldsymbol{\rho}+\boldsymbol{\rho}^\prime)\tilde{u}^*_{\ell_{\mathrm{pr}},p_{\mathrm{pr}}}(\boldsymbol{\rho})\tilde{u}^*_{\ell_{\mathrm{s}},p_{\mathrm{s}}}(\boldsymbol{\rho}^\prime).
\end{align}
If an atomic cloud is considered, and the medium spread function $N(\mathbf{r})$ is included in the calculations, the distribution of coefficients $\{a_{m,n}\}$ is modified, as we would have $\widetilde{\mathrm{V}}=\mathcal{F}[N\mathcal{V}^2_\mathrm{p}]$. In addition, the factor $e^{-4z^2/\mathcal{L}^2}$ must be included in the $z$ integral of the Eq. (\ref{eq:CT_amn}), with the integration limits being extended to infinity.
In Appendix \ref{Gaussian_Pump} we quantify the effect of a transverse size of the interaction medium comparable to the waist of the pump beam on the coefficients $\{a_{m,n}\}$ in the case of a Gaussian pump.


\subsection{Explicit formula for the biphoton coincidence amplitude in the momentum representation}
\label{appendix_TT}

To obtain an explicit formula for the biphoton coincidence amplitudes in transverse momentum space, $\widetilde{C}^{\ell_\mathrm{pr},\ell_\mathrm{s}}_{p_\mathrm{pr},p_\mathrm{s}}$, Eq. (\ref{eq:CT_amn}), we follow very closely the steps in Ref. \cite{baghdasaryan2022generalized} for the calculation of the corresponding quantity in PDC.
First, we need to evaluate $T^{\ell_\mathrm{pr},\ell_\mathrm{s},m}_{p_\mathrm{pr},p_\mathrm{s},n}(z)$, Eq. (\ref{eq:T_momentum}), which we rewrite as:
\begin{align}
    T^{\ell_{\mathrm{pr}},\ell_{\mathrm{s}},m}_{p_{\mathrm{pr}},p_{\mathrm{s}},n}(z)&=\frac{w^3_0}{8}\frac{\mathcal{N}_{\ell_{\mathrm{pr}},p_{\mathrm{pr}}}\mathcal{N}_{\ell_{\mathrm{s}},p_{\mathrm{s}}}\mathcal{N}_{m,n}e^{i\Delta_N}}{(w_0/\sqrt{2})^{-|\ell_{\mathrm{pr}}|-|\ell_{\mathrm{s}}|-|m|}} \nonumber
    \\
    &\times\int {{d}}\rho{{d}}\rho^\prime {{d}}\varphi{{d}}\varphi^\prime \mathcal{W}(\rho,\rho^\prime,\varphi,\varphi^\prime,z),
\end{align}
where $\Delta_N=N_{m,n}-N_{\ell_{\mathrm{pr}},p_{\mathrm{pr}}}-N_{\ell_{\mathrm{s}},p_{\mathrm{s}}}$ is the difference in total mode order, and the function $\mathcal{W}$ is
\begin{align}
    &\mathcal{W}(\rho,\rho^\prime,\varphi,\varphi^\prime,z)\nonumber
    \\
    &=\rho^{|\ell_{\mathrm{pr}}|+1}\rho^{\prime|\ell_{\mathrm{s}}|+1} [\rho^2+\rho^{\prime2}+2\rho\rho^\prime\cos(\varphi-\varphi^\prime)]^{\frac{|m|}{2}}\nonumber
    \\
    &\times L_{p_{\mathrm{pr}}}^{|\ell_{\mathrm{pr}}|}\left(\frac{\rho^2w^2_0}{2}\right)L_{p_{\mathrm{s}}}^{|\ell_{\mathrm{s}}|}\left(\frac{\rho^{\prime2}w^2_0}{2}\right) L_n^{|m|}\left(\frac{\rho^2_+w^2_0}{2}\right) \nonumber
    \\
    &\times e^{-i\frac{z}{4k}[\rho^2+\rho^{\prime2}-2\rho\rho^\prime\cos(\varphi-\varphi^\prime)]}e^{-\frac{w^2_0}{4}[\rho^2+\rho^{\prime2}+2\rho\rho^\prime\cos(\varphi-\varphi^\prime)]}\nonumber
    \\
    &\times e^{-\rho^2w^2_0/4}e^{-\rho^{\prime 2}w^2_0/4}e^{im\varphi_+}e^{-i(\ell_{\mathrm{pr}}\varphi+\ell_{\mathrm{s}}\varphi^\prime)},
\end{align}
with $(\rho_+,\varphi_+)=\boldsymbol{\rho}+\boldsymbol{\rho}^\prime$.
We now use the formula for the associated Laguerre polynomials,
\begin{align}
    L^{\alpha}_n(x)&=\sum^n_{k=0}\frac{(-1)^k(n+\alpha)!}{k!\,(\alpha+k)!\,(n-k)!} x^k=\sum^n_{k=0} b^{n,\alpha}_k x^k,
\end{align}
in conjunction with
the binomial expansion to write:
\begin{align}
    [\rho^2&+\rho^{\prime2}+2\rho\rho^\prime\cos(\varphi-\varphi^\prime)]^{h} \nonumber
    \\
    &=\sum^{h}_{u=0}\binom{h}{u}(\rho^2+\rho^{\prime2})^{h-u}[2\rho\rho^\prime\cos(\varphi-\varphi^\prime)]^u.
\end{align}
Also, we can express the term involving the polar angle $\varphi_+$ (the angle of the transverse momentum sum $\boldsymbol{\rho}+\boldsymbol{\rho}^\prime$) as:
\begin{align}
    e^{i m\varphi_+} &= (\cos\varphi_+ + i\sin\varphi_+)^m, \nonumber
    \\
    &= \frac{e^{im\varphi}}{\rho^m_+}(\rho+\rho^\prime e^{i(\varphi^\prime-\varphi)})^m.
\end{align}
We apply the binomial formula once again to the terms $(\rho^2+\rho^{\prime2})^{h-u}$ and $(\rho+\rho^\prime e^{i(\varphi^\prime-\varphi)})^m$, and also to the $[\cos(\varphi-\varphi^\prime)]^u=[\frac{1}{2}(e^{i(\varphi-\varphi^\prime)}+e^{-i(\varphi-\varphi^\prime)})]^u$ term.
After these steps, the azimuthal integral over the angle $\varphi$ can be evaluated using:
\begin{align}
    \frac{1}{2\pi}\int^{2\pi}_0e^{in\varphi\pm ix\cos(\varphi-\varphi^\prime)}{{d}}\varphi=(\pm i)^n e^{in\varphi^\prime}J_n(x),
\end{align}
where the Bessel function $J_\nu(x)$ is given by:
\begin{align}
    J_\nu(x)&=\sum^\infty_{k=0}\frac{(-1)^k2^{-2k-\nu}}{k!\,\Gamma(k+\nu+1)}x^{2k+\nu}, \nonumber
    \\
    &=\sum^{\infty}_{k=0} \tau_{k,\nu} x^{2k+\nu}.
\end{align}
Then, with $a=m-v-\ell_{\mathrm{pr}}+u-2d$, $c=v-\ell_{\mathrm{s}}-u+2d$, we have:
\begin{align}
    \int{{d}}\varphi{{d}}\varphi^\prime e^{ic\varphi^\prime}&e^{ia\varphi-i\beta(z)\rho\rho^\prime\cos(\varphi-\varphi^\prime)} \nonumber
    \\
    &=4\pi^2\delta_{a,-c}\,(-i)^{a}{J}_{a}\left(\beta(z)\rho\rho^\prime\right).
\end{align}
Note that $\delta_{a,-c}$ implies $\delta_{m,\ell_{\mathrm{pr}}+\ell_{\mathrm{s}}}$, thus imposing the expected OAM conservation condition in the process as $m=\ell_{\mathrm{pr}}+\ell_{\mathrm{s}}$ for all values of the pump topological charge components.
As a last step, we have to evaluate Gaussian integrals on the variables $\rho,\rho^\prime$ in the form
\begin{align}
    \int^\infty_0 \rho^n e^{-\alpha\rho^2}{{d}}\rho =\frac{\Gamma\left(\frac{n+1}{2}\right)}{2\alpha^{\frac{n+1}{2}}}.
\end{align}
In doing so, we finally obtain:
\begin{align} \label{eq:coincidence_C_q}
    T&^{\ell_{\mathrm{pr}},\ell_{\mathrm{s}},m}_{p_{\mathrm{pr}},p_{\mathrm{s}},n}(z)=\delta_{m,\ell_{\mathrm{pr}}+\ell_{\mathrm{s}}}\frac{\pi^2 w^3_0}{8}\frac{\mathcal{N}_{\ell_{\mathrm{pr}},p_{\mathrm{pr}}}\mathcal{N}_{\ell_{\mathrm{s}},p_{\mathrm{s}}}\mathcal{N}_{m,n}e^{i\frac{\pi}{2}\Delta_N}}{(w_0/\sqrt{2})^{-|\ell_{\mathrm{pr}}|-|\ell_{\mathrm{s}}|-|m|}} \nonumber
    \\
    &\times\sum^{p_{\mathrm{pr}}}_{j=0}\sum^{p_{\mathrm{s}}}_{k=0}\sum^{n}_{l=0}B^{p_{\mathrm{pr}},|\ell_{\mathrm{pr}}|}_{j}B^{p_{\mathrm{s}},|\ell_{\mathrm{s}}|}_{k}B^{n,|m|}_{l}\sum^h_{u=0}\sum^m_{v=0}\binom{h}{u}\binom{m}{v}\nonumber
    \\
    &\times\sum^{h-u}_{f=0}\sum^u_{d=0}\binom{h-u}{f}\binom{u}{d} (-i)^{a}\mathcal{Y}_{\boldsymbol{j}}(z),
\end{align}
where, $B^{n,\alpha}_k=b^{n,\alpha}_k(w^2_0/2)^k$,
\begin{align}
    \mathcal{Y}_{\boldsymbol{j}}(z)&=\frac{\beta^{a}(z)}{\alpha^{1+s}(z)}\sum^\infty_{g=0} \tau_{g,a}\widetilde{\Gamma}_{A}\widetilde{\Gamma}_{B}[\eta(z)]^{2g},
\end{align}
$\boldsymbol{j}$ represents all of the indices being summed over, $\tau_{g,a}$ is the Bessel function coefficient,
\begin{align*}
    \alpha(z)&=\frac{w^2_0}{2}+i\frac{z}{4k}, \quad\quad \beta(z)=-i\frac{w^2_0}{2}-\frac{z}{2k},
    \\
    \eta(z)&=\frac{\beta(z)}{\alpha(z)}, \quad\quad\quad\quad\, \widetilde{\Gamma}_k=\Gamma\left(\frac{k+1}{2}\right),
\end{align*}
and the coefficients are given by
\begin{align*}
    h&=\frac{1}{2}(2l+|m|-m), \quad\quad a=m-v-\ell_{\mathrm{pr}}+u-2d,
    \\
    s&=1+\frac{1}{2}(|\ell_{\mathrm{pr}}|+|\ell_{\mathrm{s}}|+m)+j+h+k+a,
    \\
    A&=1+|\ell_{\mathrm{pr}}|+2(j+h-u-f+g)+u+m-v+a,
    \\
    B&=1+|\ell_{\mathrm{s}}|+2(k+g+f)+u+v+a.
\end{align*}
Although not a very convenient computation to perform, the expression we obtained allows us to bypass the need to numerically evaluate 4D integrals on the two transverse momentum coordinates $(\boldsymbol{\rho},\boldsymbol{\rho}^\prime)$.
The biphoton coincidence amplitudes $\widetilde{C}^{\ell_{\mathrm{pr}},\ell_{\mathrm{s}}}_{p_{\mathrm{pr}},p_{\mathrm{s}}}$ can then be calculated using Eq. (\ref{eq:CT_amn}).

\subsection{Back to the Gaussian pump}
\label{Gaussian_Pump}

With the formula given by Eq. (\ref{eq:coincidence_C_q}) at hand, we return to the Gaussian pump.
In this case, $\mathcal{V}_{\mathrm{p}} = u_{0,0}$, and we have $\frac{1}{2\pi}a_{m,n}=\delta_{m,0}s^{0,0,0}_{n,0,0}=\delta_{m,0}s_n$ [see Eq. (\ref{eq:amns})]. We truncate the expansion $\mathrm{V}=\mathcal{V}^2_\mathrm{p}=\sum_{n}a_{0,n}u_{0,n}$ at $n=g$ to approximate the pump function and its Fourier transform to a desired precision.
Therefore, we may write 
\begin{align}
    \mathrm{V}(\mathbf{r}_\perp)&\simeq\mathrm{V}^{(g)}(\mathbf{r}_\perp)= \sum^g_{q=0}s_q u_{0,q}(\mathbf{r}_\perp),
    \\
    \widetilde{\mathrm{V}}(\boldsymbol{\rho})&\simeq\widetilde{\mathrm{V}}^{(g)}(\boldsymbol{\rho})=\sum^g_{q=0}s_q\tilde{u}_{0,q}(\boldsymbol{\rho}).
\end{align}
The fidelity of the approximation for a given $g$ is given by
\begin{align}
    \mathcal{F}_g =\frac{\int\mathrm{V}\,[\mathrm{V}^{(g)}]^*{{d}}^2\mathbf{r}_\perp}{\int\mathrm{V}\,\mathrm{V}^*{{d}}^2\mathbf{r}_\perp}= \frac{1}{\mathcal{N}}\sum^{g}_{k=0}|s_k|^2,
\end{align}
where $\mathcal{N}=\int\mathrm{V}\,\mathrm{V}^*{{d}}^2\mathbf{r}_\perp$ is the normalization factor.
In Table \ref{table:coeffsq} we show the values of the first four coefficients $s_q$, and the associated fidelities $\mathcal{F}_q$.
In our calculations we assume that the approximation is satisfactory for $g=2$, and therefore:
\begin{align} \label{eq:CT_Gaussian}
\widetilde{C}^{\ell_{\mathrm{pr}},\ell_{\mathrm{s}}}_{p_{\mathrm{pr}},p_{\mathrm{s}}}\propto\sum^2_{q=0}s_{q} \int^{\frac{L}{2}}_{-\frac{L}{2}}T^{\ell_{\mathrm{pr}},\ell_{\mathrm{s}},0}_{p_{\mathrm{pr}},p_{\mathrm{s}},q}(z){{d}}z.
\end{align}
\begin{table}[h!]
\centering
\caption{Coefficients and fidelities of the squared Gaussian decomposition.
}
\begin{tabular}{|c||c|c|} 
\hline
Radial index $q$& Coefficient $s_q$ & Fidelity $\mathcal{F}_q$\\
\hline
\hline
0 & 0.9429 & 0.8889\\
\hline
1 & 0.3143 & 0.9876\\
\hline
2 & 0.1048 & 0.9986\\
\hline
3 & 0.0349 & 0.9999\\
\hline
\end{tabular}
\label{table:coeffsq}
\end{table}

\begin{figure}[t!]
    \centering
    \includegraphics[width=1\linewidth,trim={0cm 0.5cm 0cm 0cm}]{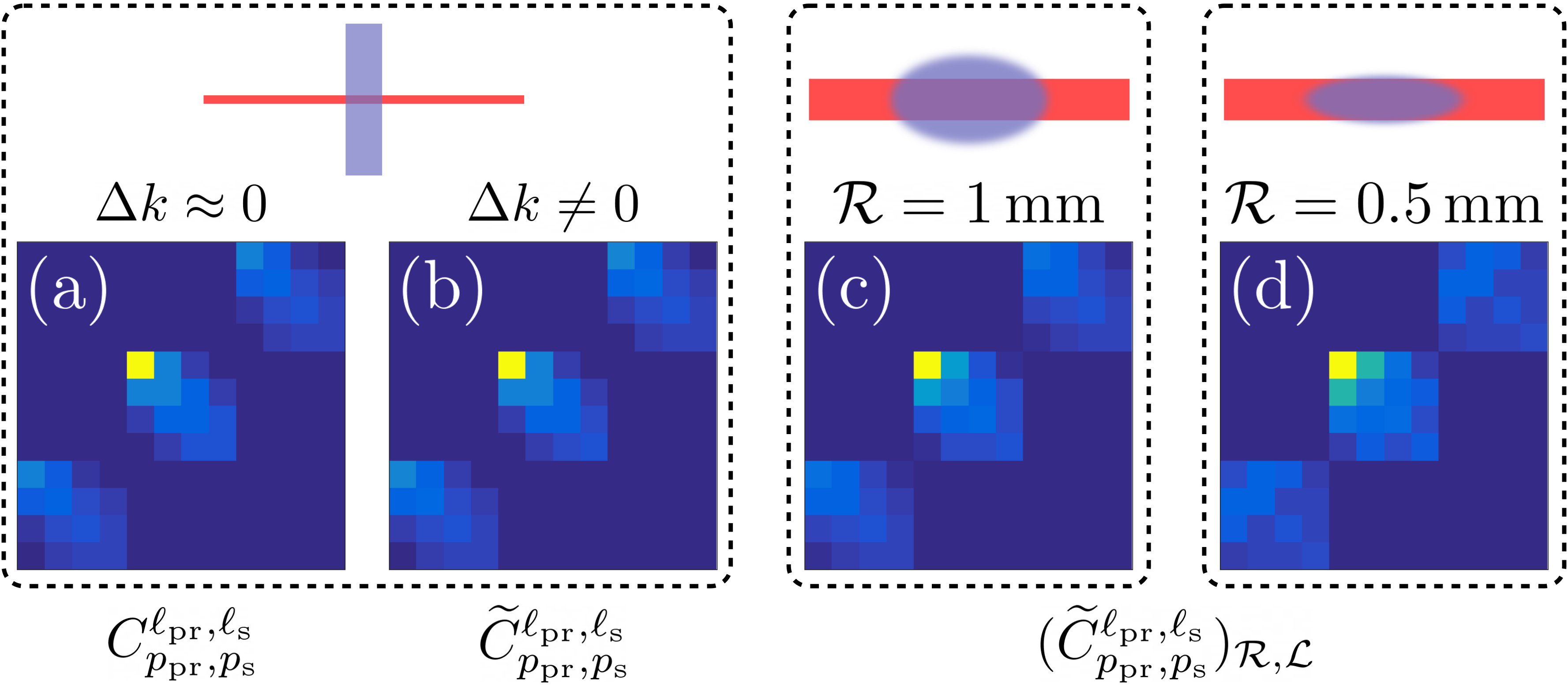}
    \caption{Biphoton coincidence amplitudes in the subspace $\mathbb{S}(1,3)$ calculated with Eq. (\ref{eq:C_A}) (a), Eq. (\ref{eq:CT_Gaussian}) (b), and Eq. (\ref{eq:C_tilde_RL}) (c),(d). In (a) and (b) the medium is uniformly distributed with length $L=3\,\mathrm{mm}$. In (c) and (d) the medium geometry is a cloud of atoms with the same longitudinal size $\mathcal{L}=3\,\mathrm{mm}$ and different radius $\mathcal{R}$. Illustrations at the top indicate the corresponding physical situations.}
    \label{fig:Medium_Geom}
\end{figure}

As mentioned, to take into account the transverse and longitudinal characteristic lengths of a cold atom cloud 
$(\mathcal{R},\mathcal{L})$, we can write
\begin{align} \label{eq:C_tilde_RL}
(\widetilde{C}^{\ell_{\mathrm{pr}},\ell_{\mathrm{s}}}_{p_{\mathrm{pr}},p_{\mathrm{s}}})&_{\mathcal{R},\mathcal{L}}\propto \sum^2_{q=0}a_{q}(\xi) \int T^{\ell_{\mathrm{pr}},\ell_{\mathrm{s}},0}_{p_{\mathrm{pr}},p_{\mathrm{s}},q}(z)e^{-\frac{4z^2}{\mathcal{L}^2}}{{d}}z,
\end{align}
where $\xi=\mathcal{R}/w_0$, $a_q(\xi)$ is the pump function expansion coefficient taking into account the cloud radius, and the integration region should accommodate all of the relevant variation of the integrand.
To quantify the effect of the geometry of the medium on the pump function, we show in Table \ref{table:coeffsq_R} the modified expansion coefficients in case $\xi=0.5,1,3$.
We see that there is a sensible modification to the coefficients when $\mathcal{R}=w_0/2,w_0$, while for $\mathcal{R}=3w_0$, the expansion is very similar to the one calculated considering a uniform transverse distribution.
\begin{table}[h!]
\centering
\caption{Coefficients of the pump function decomposition considering the geometry of an atom cloud with $\xi=0.5,1,3$.
}
\begin{tabular}{|c||c||c||c||c|} 
\hline
$q$& $a_q(\xi\rightarrow\infty)$ & $a_q(\xi=3)$ & $a_q(\xi=1)$ & $a_q(\xi=0.5)$\\
\hline
\hline
0 & 0.9429 & 0.9397 & 0.8677 & 0.7248 \\
\hline
1 & 0.3143 & 0.3215 & 0.4339 & 0.5177\\
\hline
2 & 0.1048 & 0.1100 & 0.2169 & 0.3698\\
\hline
3 & 0.0349 & 0.0376 & 0.1085 & 0.2642\\
\hline
\end{tabular}
\label{table:coeffsq_R}
\end{table}
Now, to see how the cloud geometry affects the resulting generated light-state, we show in Fig. \ref{fig:Medium_Geom} a comparison between the coincidence amplitudes calculated considering a uniform atom distribution (equivalent to a crystal), using Eq. (\ref{eq:C_A}) ($\Delta k\approx0$), Eq. (\ref{eq:CT_Gaussian}) ($\Delta k\neq0$), and those calculated with Eq. (\ref{eq:C_tilde_RL}), when we consider different values for $\mathcal{R},\mathcal{L}$.
In all situations, we considered $w_0=1\,\mathrm{mm}$.
Indeed we can see a more clear perturbation to the mode distribution for $\mathcal{R}=0.5w_0$.
The influence of the medium extension $\mathcal{L}$ is not as relevant as that of the radius $\mathcal{R}$ in this case, since $\mathcal{L}/z_R\ll1$ for $w_0=1\,\mathrm{mm}$.


\newpage

\twocolumngrid
\bibliography{apssamp}


\end{document}